\documentclass[aps, prd, floatfix, nofootinbib, superscriptaddress, twocolumn]{revtex4-1}

\usepackage{latexsym}
\usepackage{amsmath}
\usepackage{amssymb}
\usepackage{amsfonts}

\usepackage[mathscr,scaled=1.15]{urwchancal}
\DeclareFontFamily{OT1}{pzc}{}
\DeclareFontShape{OT1}{pzc}{m}{it}%
{<-> s * [1.15] pzcmi7t}{}
\DeclareMathAlphabet{\mathpzc}{OT1}{pzc}{m}{it}

\usepackage{color}

\usepackage{supertabular}
\usepackage{placeins}
\usepackage{epsfig}
\usepackage{graphicx}

\definecolor{purple}{rgb}{0.5,0,0.5}
\definecolor{blue}{rgb}{0.0,0,0.9}
\definecolor{prdblue}{rgb}{0.133,0.118,0.498}
\usepackage[colorlinks=true, pdfstartview=FitV, linkcolor=prdblue, citecolor= prdblue, urlcolor=prdblue]{hyperref}

\begin{document}


\title{$\,$\\[-7ex]\hspace*{\fill}{\normalsize{\sf\emph{Preprint no}. NJU-INP 017/20}}\\[1ex]
Nucleon elastic form factors at accessible large spacelike momenta}

\author{Zhu-Fang Cui}
\email[]{phycui@nju.edu.cn}
\affiliation{School of Physics, Nanjing University, Nanjing, Jiangsu 210093, China}
\affiliation{Institute for Nonperturbative Physics, Nanjing University, Nanjing, Jiangsu 210093, China}
\author{Chen Chen}
\email[]{Chen.Chen@theo.physik.uni-giessen.de}
\affiliation{Institut f\"ur Theoretische Physik, Justus-Liebig-Universit\"at Gie{\ss}en, D-35392 Gie{\ss}en, Germany}
\author{Daniele Binosi}
\email{binosi@ectstar.eu}
\affiliation{European Centre for Theoretical Studies in Nuclear Physics
and Related Areas; Villa Tambosi, Strada delle Tabarelle 286, I-38123 Villazzano (TN), Italy}
\author{Feliciano De Soto}
\email[]{fcsotbor@upo.es}
\affiliation{Dpto. Sistemas F\'isicos, Qu\'imicos y Naturales, Univ.\ Pablo de Olavide, E-41013 Sevilla, Spain}
\author{Craig D.~Roberts}
\email[]{cdroberts@nju.edu.cn}
\affiliation{School of Physics, Nanjing University, Nanjing, Jiangsu 210093, China}
\affiliation{Institute for Nonperturbative Physics, Nanjing University, Nanjing, Jiangsu 210093, China}
\author{Jos\'e Rodr\'{\i}guez-Quintero}
\email[]{jose.rodriguez@dfaie.uhu.es}
\affiliation{Department of Integrated Sciences and Center for Advanced Studies in Physics, Mathematics and Computation;
University of Huelva, E-21071 Huelva; Spain.}
\author{Sebastian M.~Schmidt}
\email[]{s.schmidt@hzdr.de}
\affiliation{RWTH Aachen University, III. Physikalisches Institut B, Aachen D-52074, Germany}
\affiliation{Helmholtz-Zentrum Dresden-Rossendorf, Dresden D-01314, Germany}
\author{Jorge Segovia}
\email[]{jsegovia@upo.es}
\affiliation{Dpto.\ Sistemas F\'isicos, Qu\'imicos y Naturales, Univ.\ Pablo de Olavide, E-41013 Sevilla, Spain}
\affiliation{Institute for Nonperturbative Physics, Nanjing University, Nanjing, Jiangsu 210093, China}

\date{25 March 2020}

\begin{abstract}
A Poincar\'e-covariant quark+diquark Faddeev equation is used to compute nucleon elastic form factors on $0\leq Q^2\leq 18 \,m_N^2$ ($m_N$ is the nucleon mass) and elucidate their role as probes of emergent hadronic mass in the Standard Model.  The calculations expose features of the form factors that can be tested in new generation experiments at existing facilities, \emph{e.g}.\ a zero in $G_E^p/G_M^p$; a maximum in $G_E^n/G_M^n$; and a zero in the proton's $d$-quark Dirac form factor, $F_1^d$.  Additionally, examination of the associated light-front-transverse number and anomalous magnetisation densities reveals, \emph{inter alia}: a marked excess of valence $u$-quarks in the neighbourhood of the proton's centre of transverse momentum; and that the valence $d$-quark is markedly more active magnetically than either of the valence $u$-quarks.  The calculations and analysis also reveal other aspects of nucleon structure that could be tested with a high-luminosity accelerator capable of delivering higher beam energies than are currently available.
\end{abstract}

\maketitle


\section{Introduction}
\label{Introduction}
Ever since it was found that the proton and neutron are composite systems \cite{Hofstadter:1955ae}, their elastic electromagnetic form factors have been the focus of extensive programmes in both experiment and theory.  In the context of quantum chromodynamics (QCD), \emph{i.e}.\ strong interactions within the Standard Model, this is because they can provide insights into key features of nucleon structure, such as the role played by emergent hadronic mass (EHM) in determining the proton's size and fixing both the location and rate of the transition between the strong and perturbative domains of QCD.

Experiments completed during the past twenty years have had a big impact.  For instance, they have revealed that despite its simple valence-quark content, the internal structure of the proton is very complex, with marked differences between the distributions of total charge and magnetisation \cite{Jones:1999rz,Gayou:2001qd, Punjabi:2005wq, Puckett:2010ac, Puckett:2011xg, Puckett:2017flj} and also between the distributions carried by the different quark flavours \cite{Cates:2011pz, Wojtsekhowski:2020tlo}.  The goals now are to connect these and related observations for the neutron with properties of QCD; and to make robust predictions at large-$Q^2$.  Such predictions are particularly important because new experiments are approved at Jefferson Lab (JLab) that will acquire data at unprecedented photon virtualities, \emph{e.g}.\ \cite{Gilfoyle:2018xsa, Wojtsekhowski:2020tlo}:
proton electric form factor to $Q^2= 12\,$GeV$^2$ \cite{E12-07-109};
proton magnetic form factor to $Q^2= 15.5\,$GeV$^2$ \cite{E12-07-108};
neutron electric form factor to $10.2\,$GeV$^2$ \cite{E12-09-016};
and neutron magnetic form factor to $13.5\,$GeV$^2$ \cite{E12-09-019}.
In principle, given that JLab beam energy allows for measurements reaching to $Q^2=18\,$GeV$^2$, it might be possible in some of these cases to obtain data at higher momentum transfers; but here the issue of achievable precision needs to be explored \cite{BogdanPrivate2020}.

An \emph{ab initio} approach to delivering QCD-connected predictions for nucleon form factors is provided by the numerical simulation of lattice-regularised QCD (lQCD).  In reaching high-$Q^2$ using this framework, a number of obstacles must be overcome, \emph{e.g}.\ one should use, \emph{inter alia}: a large lattice volume to accommodate physically light quarks; small lattice spacing; and high statistics to offset a decaying signal-to-noise ratio as form factors drop rapidly with increasing $Q^2$.  To master these challenges, new algorithms are being tested and preliminary results are available \cite{Chambers:2017tuf, Kallidonis:2018cas}.

On the other hand, phenomenological models have long been used to draw insights from available data and make projections to aid further developments in experiment \cite{Arrington:2006zm, Perdrisat:2006hj, Denig:2012by, Punjabi:2015bba}.  Here, in pushing beyond $Q^2 = m_N^2$, where $m_N$ is the nucleon mass, a Poincar\'e-covariant framework is very useful; and an approach grounded in quantum field theory is crucial if QCD features, such as momentum-dependent dressed-quark mass-functions \cite{Bhagwat:2003vw, Bowman:2005vx, Bhagwat:2006tu} and anomalous magnetic moments \cite{Singh:1985sg, Bicudo:1998qb, Chang:2010hb,  Chang:2011ei}, both signatures of EHM, are to be incorporated and their impacts expressed.

Continuum Schwinger-function methods (CSMs) meet these requirements.  Formulated optimally, they provide a systematic, symmetry-preserving approach to solving problems in QCD \cite{Roberts:2015lja, Horn:2016rip, Binosi:2016rxz, Eichmann:2016yit, Burkert:2019bhp, Fischer:2018sdj}.  Moreover, where fair comparisons can be drawn, predictions from such continuum analyses match those obtained using lQCD; hence, the approaches are complementary and there is real synergistic potential.

Some applications of CSMs to the computation of nucleon form factors are summarised in Ref.\,\cite{Eichmann:2016yit}.  This body of work follows earlier calculations of the pion form factor \cite{Maris:2000sk}, \emph{viz}.\ it assumes a form for the quark-antiquark interaction, solves the Dyson-Schwinger equation (DSE) for each $n$-point function that appears in a rainbow-ladder (leading-order) approximation to the nucleon form factors, then computes the form factors using these numerically determined inputs.  However, reliable predictions on $x=Q^2/m_N^2 \gtrsim 8$ are not yet available owing to limitations of the numerical algorithms employed \cite{Eichmann:2011vu}.

Analogous algorithms were employed for mesons in Ref.\,\cite{Maris:2000sk} with the same outcome, \emph{i.e}.\ a curtailed $Q^2$-domain for the computed form factor.  A solution to this problem is described in Ref.\,\cite{Chang:2013nia}.  Based on the use of perturbation theory integral representations for all the Schwinger functions needed in a given form factor calculation, it has since been used widely for mesons \cite{Shi:2015esa, Raya:2015gva, Raya:2016yuj, Gao:2017mmp, Ding:2018xwy}.  Applications to baryons are beginning.

Predictions for the large-$Q^2$ behaviour of nucleon form factors have been delivered using a fully dynamical quark+diquark reduction of the Poincar\'e-covariant three-body bound-state problem in relativistic quantum field theory \cite{Segovia:2014aza}.  Herein, we reconsider those calculations because questions can also be asked about the algorithms used to compute the form factors on $x=Q^2/m_N^2 \gtrsim 10$.  Namely, the interaction current involves two-loop diagrams.  Monte-Carlo methods are required for their evaluation; and with any finite number of samples, such methods are imprecise when the answer is a small number, as is the case with form factors at large photon virtuality, and not all contributions are of the same sign.

The manuscript is composed as follows.  The quark+diquark Faddeev equation for the nucleon and its inputs are sketched in Sec.\,\ref{SecCurrents}, which also incorporates a brief description of the electromagnetic interaction current.  Our approach to calculating the elastic form factors is detailed in Sec.\,\ref{CalcFFs}, including an explanation of a statistical implementation of the Schlessinger point method (SPM) for the interpolation and extrapolation of smooth functions \cite{Schlessinger:1966zz, PhysRev.167.1411, Tripolt:2016cya, Chen:2018nsg, Binosi:2018rht, Binosi:2019ecz}.   Predictions for all nucleon elastic form factors are reported and discussed in Sec.\,\ref{SecElastic}.  Section\,\ref{SecFlavour} canvasses the flavour separation of these form factors and includes an analysis of the associated light-front-transverse valence-quark number and anomalous magnetisation densities.  Section~\ref{Epilogue} is a summary and outlook.

\section{Nucleon Structure and Electromagnetic Currents}
\label{SecCurrents}
We employ the description of nucleon structure and electromagnetic currents detailed in Ref.\,\cite{Segovia:2014aza}; hence, only provide a brief recapitulation herein.

\begin{figure}[t]
\centerline{%
\includegraphics[clip, width=0.45\textwidth]{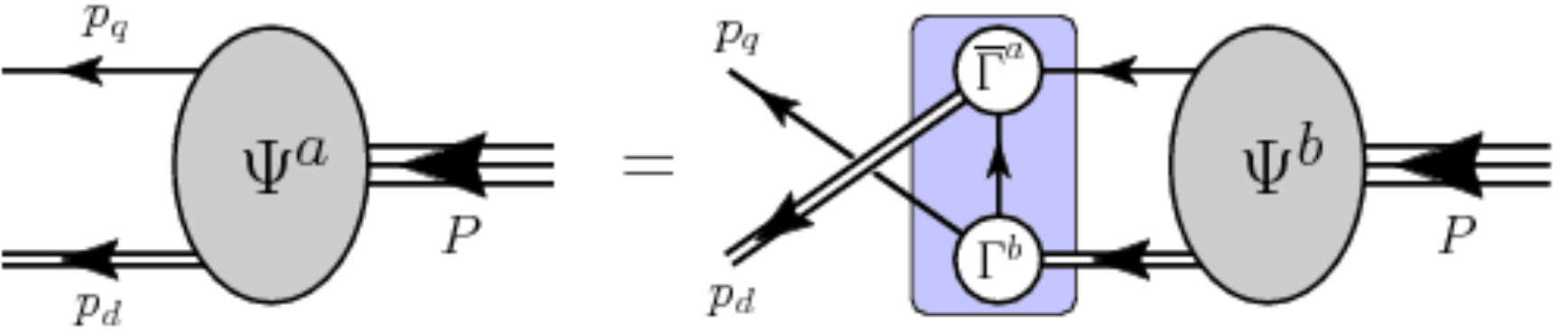}}
\caption{\label{figFaddeev}
Nucleon = quark+diquark Faddeev equation.  This is a linear integral equation for the Poincar\'e-covariant matrix-valued function $\Psi$, the Faddeev amplitude for a state with total momentum $P= p_q + p_d$.  It describes the relative momentum correlation between the dressed-quarks and -diquarks.  Legend. Shaded rectangle -- kernel of the Faddeev equation; \emph{single line} -- dressed-quark propagator; $\Gamma$ -- diquark correlation amplitude; and \emph{double line} -- diquark propagator.
Ground-state nucleons ($n$ - neutron, $p$ - proton) contain both isoscalar-scalar diquarks, $[ud]\in (n,p)$, and isovector-pseudovector diquarks $\{dd\}\in n$, $\{ud\}\in (n,p)$, $\{uu\}\in p$.
}
\end{figure}

To begin, nucleons are described by the quark+diquark Faddeev equation introduced in Refs.\,\cite{Cahill:1988dx, Reinhardt:1989rw, Efimov:1990uz} and depicted in Fig.\,\ref{figFaddeev}.  Evidence supporting the presence of diquark correlations in baryons is accumulating \cite{Eichmann:2009qa, Cates:2011pz, Wojtsekhowski:2020tlo, Eichmann:2011aa, Segovia:2013rca, Roberts:2013mja, Segovia:2014aza, Segovia:2015hra, Segovia:2015ufa, Eichmann:2016jqx, Segovia:2016zyc, Eichmann:2016hgl, Eichmann:2016nsu, Lu:2017cln, Chen:2017pse, Mezrag:2017znp, Roberts:2018hpf, Chen:2018nsg, Burkert:2019bhp, Chen:2019fzn, Lu:2019bjs}, with confirmation also found using lQCD \cite{Alexandrou:2006cq, DeGrand:2007vu, Babich:2007ah, Bi:2015ifa}.

Notably, the diquarks in Fig.\,\ref{figFaddeev} are fully dynamical: they appear in a Faddeev kernel -- shaded domain -- which requires their continual breakup and reformation.  Hence, they are a marked evolution of the pointlike, static diquarks introduced long ago \cite{Anselmino:1992vg} as an attempt to solve the so-called ``missing resonance'' problem \cite{Aznauryan:2011ub}.  Matching indications from lQCD \cite{Edwards:2011jj}, baryon spectra generated by the Faddeev equation in Fig.\,\ref{figFaddeev} are far richer than those obtained using any two-body model.

In solving the Faddeev equation, we use the following diquark masses (in GeV):
\begin{equation}
m_{[ud]} = 0.80\,,\;
m_{\{uu\}} = m_{\{ud\}} = m_{\{dd\}} = 0.89\,;
\end{equation}
and light-quarks characterised by a Euclidean constituent mass $M_{u,d}^E = 0.33\,$GeV.  The associated propagators and additional details concerning the Faddeev kernel are presented in Appendices A.1, A.2 in Ref.\,\cite{Segovia:2014aza}.  Importantly, \emph{e.g}.\ the light-quark mass function, illustrated in Ref.\,\cite[Fig.\,A.1]{Chen:2017pse}, agrees well with that obtained in modern gap equation studies \cite{Chang:2010hb, Chang:2011ei}.

With these inputs, one obtains the nucleon Faddeev amplitude, $\Psi$, and mass $m_N = 1.18\,$GeV.  This value is deliberately large because Fig.\,\ref{figFaddeev} describes the nucleon's \emph{dressed-quark core}.  The complete nucleon is obtained by including resonant contributions to the Faddeev kernel, \emph{i.e}.\ a meson cloud.  Such effects produce a physical nucleon whose mass is $\approx 0.2$\,GeV lower than that of the core \cite{Ishii:1998tw, Hecht:2002ej, Sanchis-Alepuz:2014wea}.  Their impact on baryon structure and form factors can very effectively be incorporated using dynamical coupled-channels models \cite{Aznauryan:2012ba}; and regarding nucleon-related form factors, the effects are restricted to $Q^2\lesssim 2\,m_N^2$, \emph{e.g}.\ Refs.\,\cite{Eichmann:2011aa, Segovia:2013rca, Segovia:2015hra, Chen:2018nsg, Burkert:2019bhp, Lu:2019bjs}.

\begin{figure}[!t]
\centerline{\includegraphics[clip,width=1.0\linewidth]{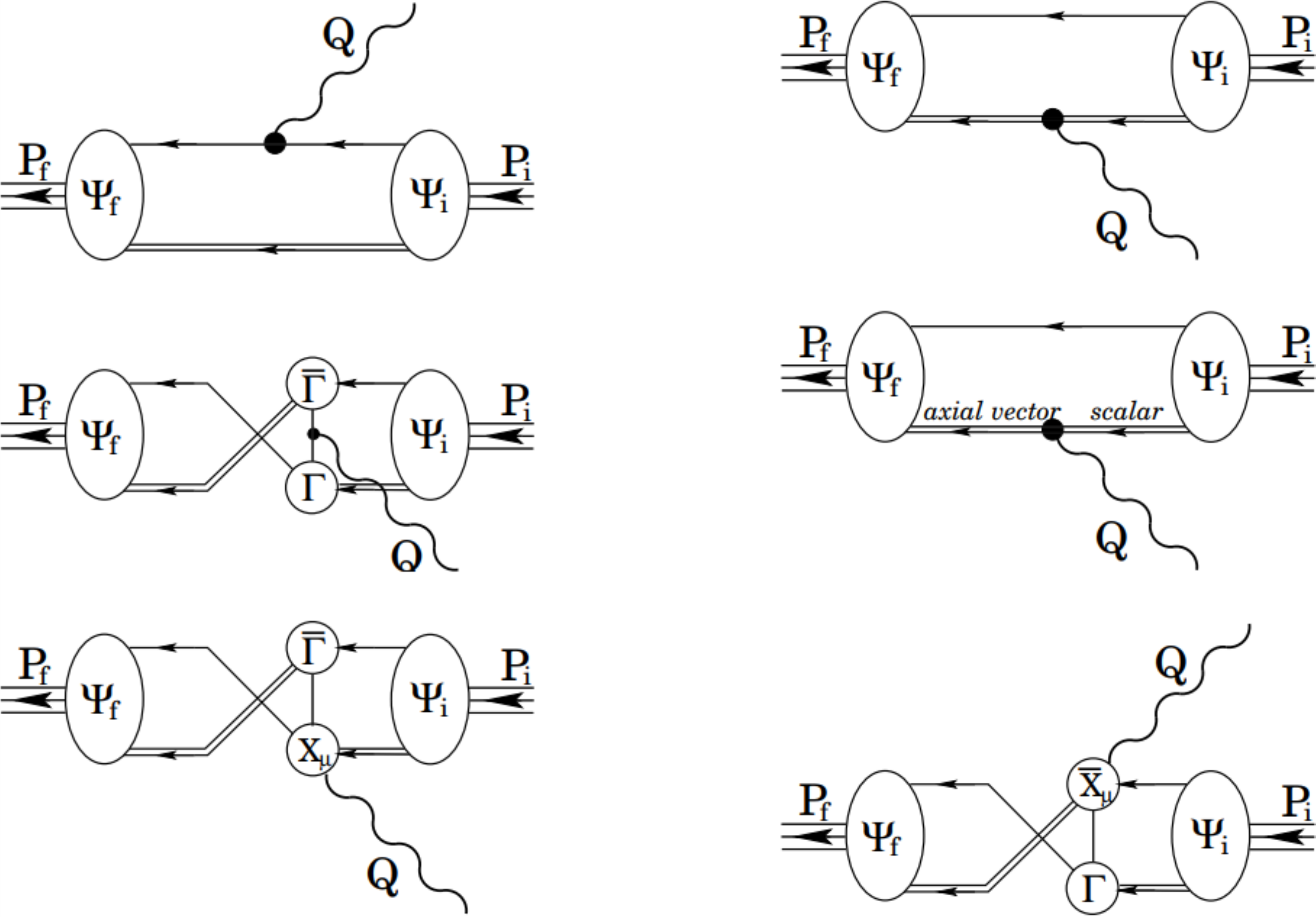}}
\caption{\label{vertexB}
Vertex that ensures a conserved current for on-shell nucleons that are described by the Faddeev amplitudes obtained from the equation illustrated in Fig.\,\ref{figFaddeev}: \emph{single line}, dressed-quark propagator; \emph{undulating line}, photon; $\Gamma$,  diquark correlation amplitude; and \emph{double line}, diquark propagator.  Diagram~1 is the top-left image; the top-right is Diagram~2;
and so on, with Diagram~6 being the bottom-right image.
Diagrams~3, 5, 6 are eight-dimensional integrals.  Monte-Carlo methods are required for their evaluation.
(Details are provided in Ref.\,\cite[Appendix~C]{Segovia:2014aza})
}
\end{figure}

The nucleon current is specified by two form factors, Dirac and Pauli:
\begin{equation}
\bar u(P_f)\big[ \gamma_\mu F_{1}(Q^2)+\frac{1}{2 m_N} \sigma_{\mu\nu} Q_\nu F_{2}(Q^2)\big] u(P_i)\,,
\label{NRcurrents}
\end{equation}
where: $u$, $\bar u$ are, respectively, Dirac spinors describing the incoming/outgoing nucleon, with four-momenta $P_{i,f}$, $P_{i,f}^2=-m_N^2$, $Q=P_f-P_i$.  The Sachs electric and magnetic form factors are ($x=Q^2/M_N^2$):
\begin{equation}
\label{GEGM}
G_E(x) = F_1(x) - \frac{x}{4} F_2(x)\,,
G_M(x) = F_1(x) + F_2(x)\,.
\end{equation}

A vertex sufficient to express the interaction of a photon with a nucleon generated by the Faddeev equation in Fig.\,\ref{figFaddeev} is described elsewhere \cite{Oettel:1999gc, Segovia:2014aza}.  It is a sum of six terms, depicted in Fig.\,\ref{vertexB}, with the photon separately probing the quarks and diquarks in various ways.  Hence, diverse features of quark dressing, quark-quark correlations, and their relative wave functions play a role in determining the form factors.

\section{Form Factor Interpolation and Extrapolation}
\label{CalcFFs}
All elastic form factors were computed herein using the algorithm described in Ref.\,\cite[Appendix~D]{Segovia:2014aza}.  In solving the Faddeev equation, ten Chebyshev polynomials were employed to express the $k\cdot P$ dependence of the Faddeev amplitude, $\Psi(k;P)$.  Regarding the integrals depicted in Fig.\,\ref{vertexB}: to compute Diagrams~1, 2, 4, we used 150 quadrature points for the momentum magnitude and 50 for both angles; and for Diagrams~3, 5, 6, $10^8$ Monte Carlo points.  These choices ensured stable results on $x\leq 9$.

On $x\gtrsim 9$, decelerating convergence of the Chebyshev expansion of $\Psi(k;P)$, \emph{i.e}.\ increasing noise and spurious oscillations, compounds inadequacies in the Monte Carlo approach to integrating functions of non-uniform sign so that a direct (brute force) approach failed to deliver precise results for the elastic form factors.

These difficulties must be surmounted in order to achieve our goal of providing precise Faddeev equation predictions extending to $x \simeq 18$, \emph{i.e}.\ the farthest reach of approved experiments at JLab.  We proceeded by capitalising on a recently developed technique that adds a powerful statistical element to the SPM \cite{Schlessinger:1966zz, PhysRev.167.1411, Tripolt:2016cya, Chen:2018nsg, Binosi:2018rht, Binosi:2019ecz}.  This approach was first exploited in Ref.\,\cite{Chen:2018nsg}, which extended the range of predictions for nucleon-to-Roper-resonance transition form factors out to $x=12$; and has since been used to analyse vector-meson elastic form factors and the domain of validity of vector meson dominance models \cite{Xu:2019ilh} and the semileptonic decays of $D_{(s)}$ mesons \cite{Yao:2019}.

To introduce the approach, suppose that one has $N$ pairs, ${\mathsf D} = \{(x_i,y_i=f(x_i))$\}, being the values of some smooth function, $f(x)$, at a given set of discrete points.  A basic SPM application constructs a continued-fraction interpolation:
\begin{equation}
F(x) = \frac{y_1}{1+\frac{a_1(x-x_1)}{{1+\frac{a_2(x-x_2)}{\vdots a_{N-1}(x-x_{N-1})}}}}
\end{equation}
in which the coefficients $\{a_i|i=1,\ldots, N-1\}$ are determined recursively and ensure $F(x_i) = f(x_i)$, $i=1\,\ldots,N$.  The SPM is related to the Pad\'e approximant; and the procedure accurately reconstructs any analytic function within a radius of convergence fixed by that one of the function's branch points which lies closest to the domain of real-axis points containing the data sample.

To give an example, suppose the target function is a monopole form factor represented by $N$ points, $0<N< \infty$, each of which lies exactly on the curve.  Then using any single point, the SPM will reproduce the monopole precisely.  If each of the set's points has some numerical error, as is common in form factor calculations, then from any single point, the SPM will deliver an analytic approximation to the form factor.  Choosing many single points at random and using the SPM with each point, then one obtains a collection of analytic approximations to the monopole whose spread measures the uncertainty inherent in the numerical calculation.  Each one of the approximations is of practically equal quality to the best least-squares fit.

\begin{figure*}[!t]
\begin{center}
\begin{tabular}{lr}
\includegraphics[clip,width=0.425\linewidth]{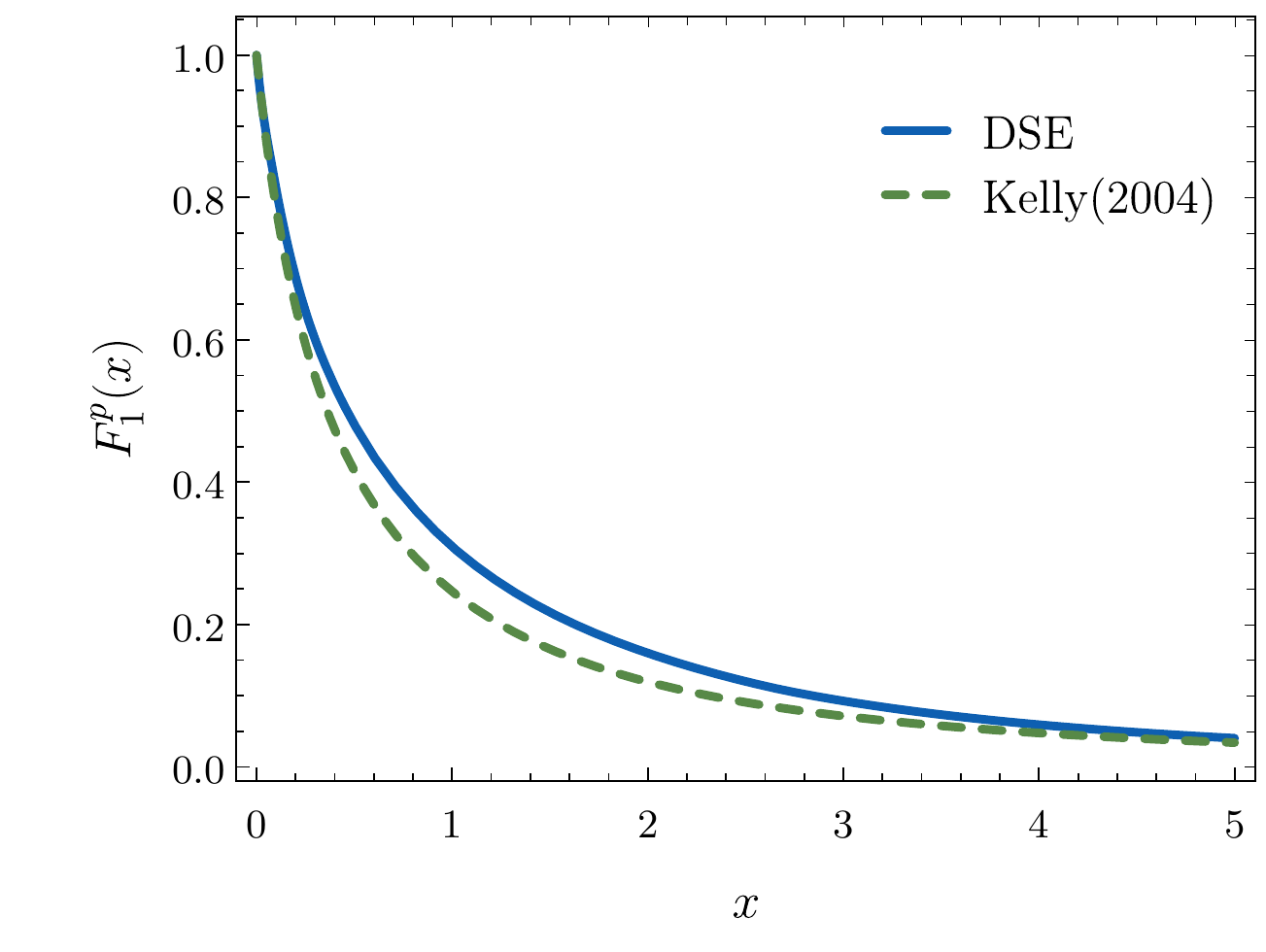}\hspace*{2ex } &
\includegraphics[clip,width=0.425\linewidth]{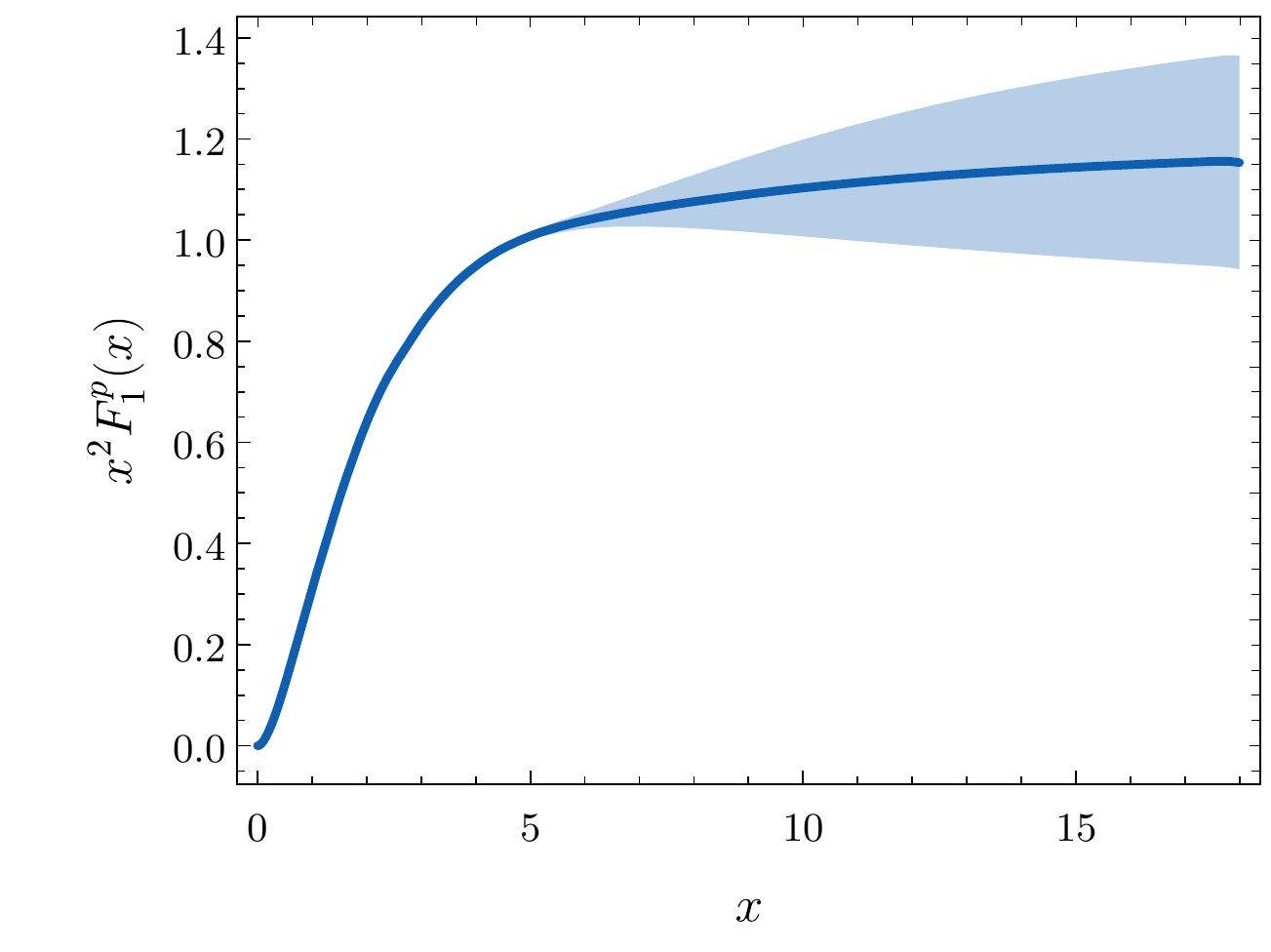}\vspace*{-0ex}
\end{tabular}
\begin{tabular}{lr}
\includegraphics[clip,width=0.425\linewidth]{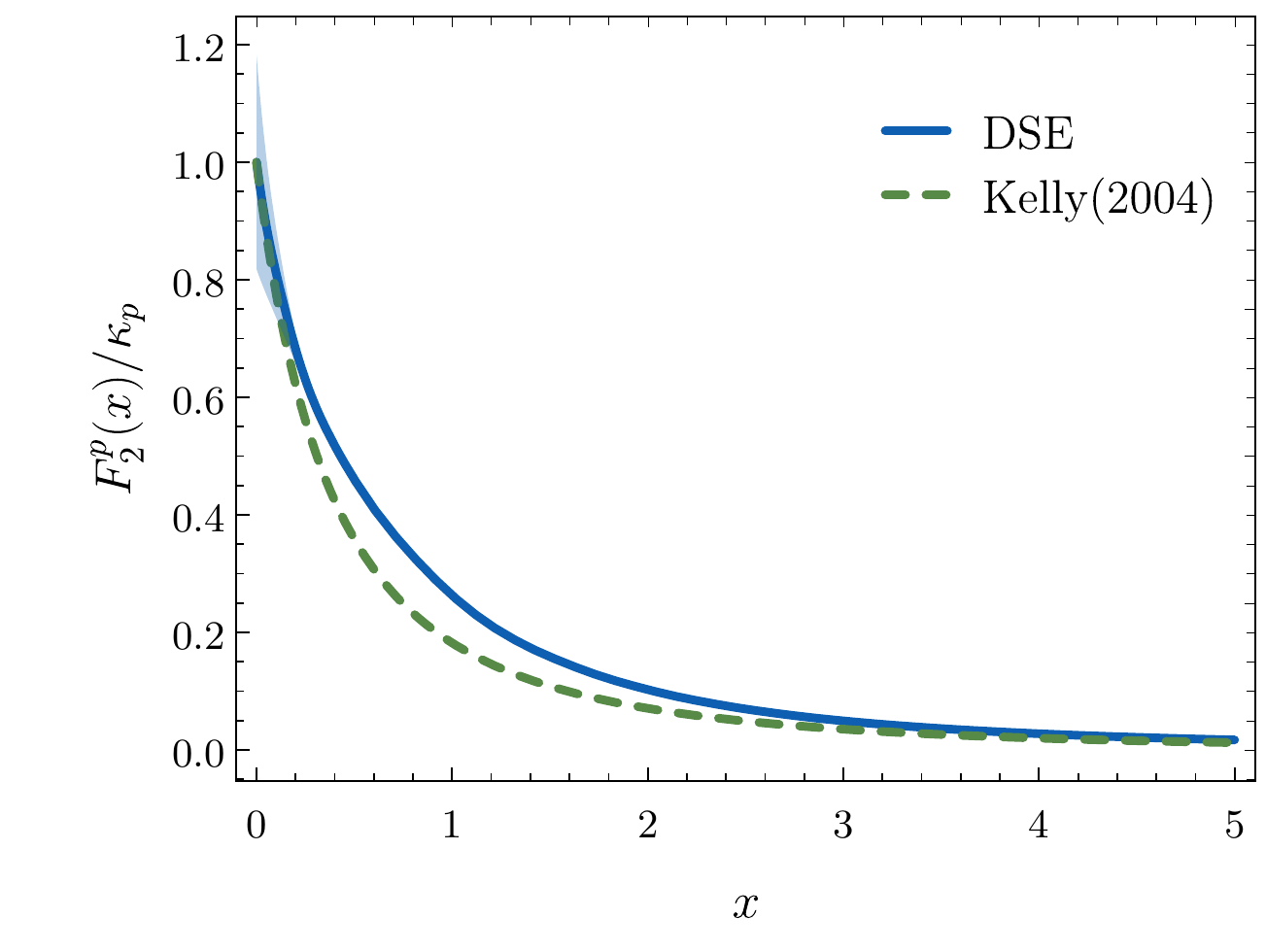}\hspace*{2ex } &
\includegraphics[clip,width=0.425\linewidth]{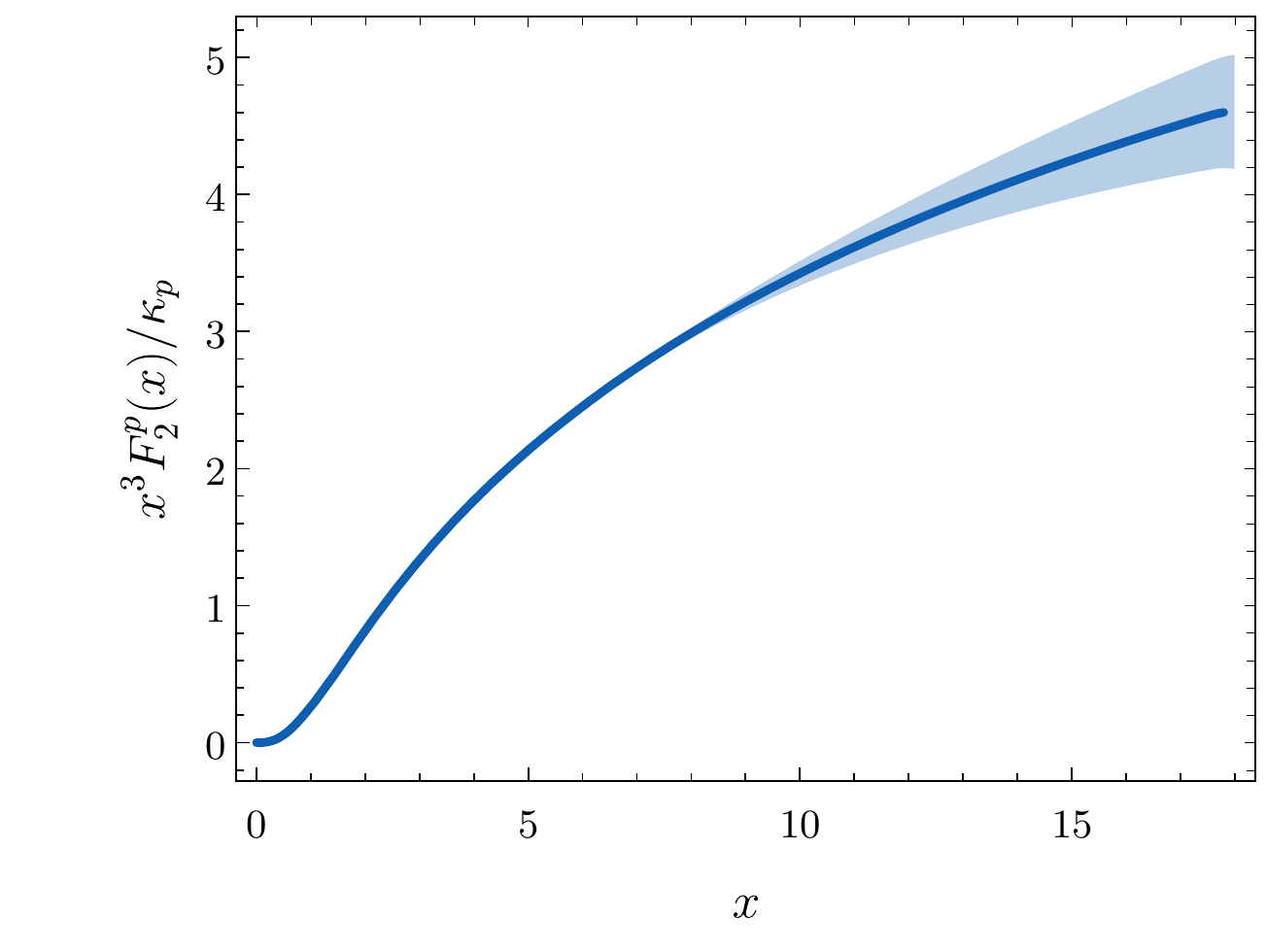}\vspace*{-1ex}
\end{tabular}
\end{center}
%
\caption{\label{FFDPp}
Proton: herein, solid blue curves.
\emph{Upper panels} --  Dirac form factor, $F_1^p(x)$, $x=Q^2/m_N^2$;
and \emph{lower panels} -- Pauli form factor, $F_2^p(x)$.   ($\kappa_p = 1.50$.)
In both cases, the right panels depict $x^s$-weighted form factors, with $s$ chosen to match the associated na\"ive scaling dimension.
The empirical fits from Ref.\,\cite{Kelly:2004hm} are displayed in the left panels (dashed green curves).
In all panels, the $1\sigma$ band for the SPM approximants is shaded in light-blue, \emph{i.e}.\ 68\% of all SPM approximants lie within the light-blue band centred on the blue curve.
}
\end{figure*}

In each of the realistic cases considered herein, we have $N$ results computed directly within the domain ${\mathsf X}= \{x\,|\,0\leq x\leq x_{\rm max}\}$.  We choose $x_{\rm max} = 4.5$; hence, the sample domain is half the size of the complete space upon which precise results are obtained using straightforward calculation.  Each of the $N$ results possesses a similar level of precision.

From this set of $N$ results, we randomly choose first one point, then two, etc., until that minimal number of points $M<N$ is reached for which the analytic approximation produced by the SPM from any randomly chosen set of $M$ points typically delivers a valid fit to the output.

The extrapolation is then defined by randomly choosing a large number, $T$, of $M$-point samples; determining the SPM approximation from each collection; applying known physical constraints, such as continuity, regularity, etc., to eliminate those functions which are unacceptable, thereby taking $T\to T_P$; and then drawing the associated extrapolation curve for each surviving approximation.
Herein, $N=500$, $M=12$, $T\approx 500\,000$, $T_P \approx 1000$.

This procedure generates a band of extrapolated curves whose collective reliability at any $x> x_{\rm max}$ is expressed by the width of the band at that point, which is itself determined by the precision of the original results on $x \leq x_{\rm max}$.
The test of the method is the precision with which it reproduces the known results computed directly on $x_{\rm max}<x<2 x_{\rm max}$.

\begin{figure*}[!t]
\begin{center}
\begin{tabular}{lr}
\includegraphics[clip,width=0.425\linewidth]{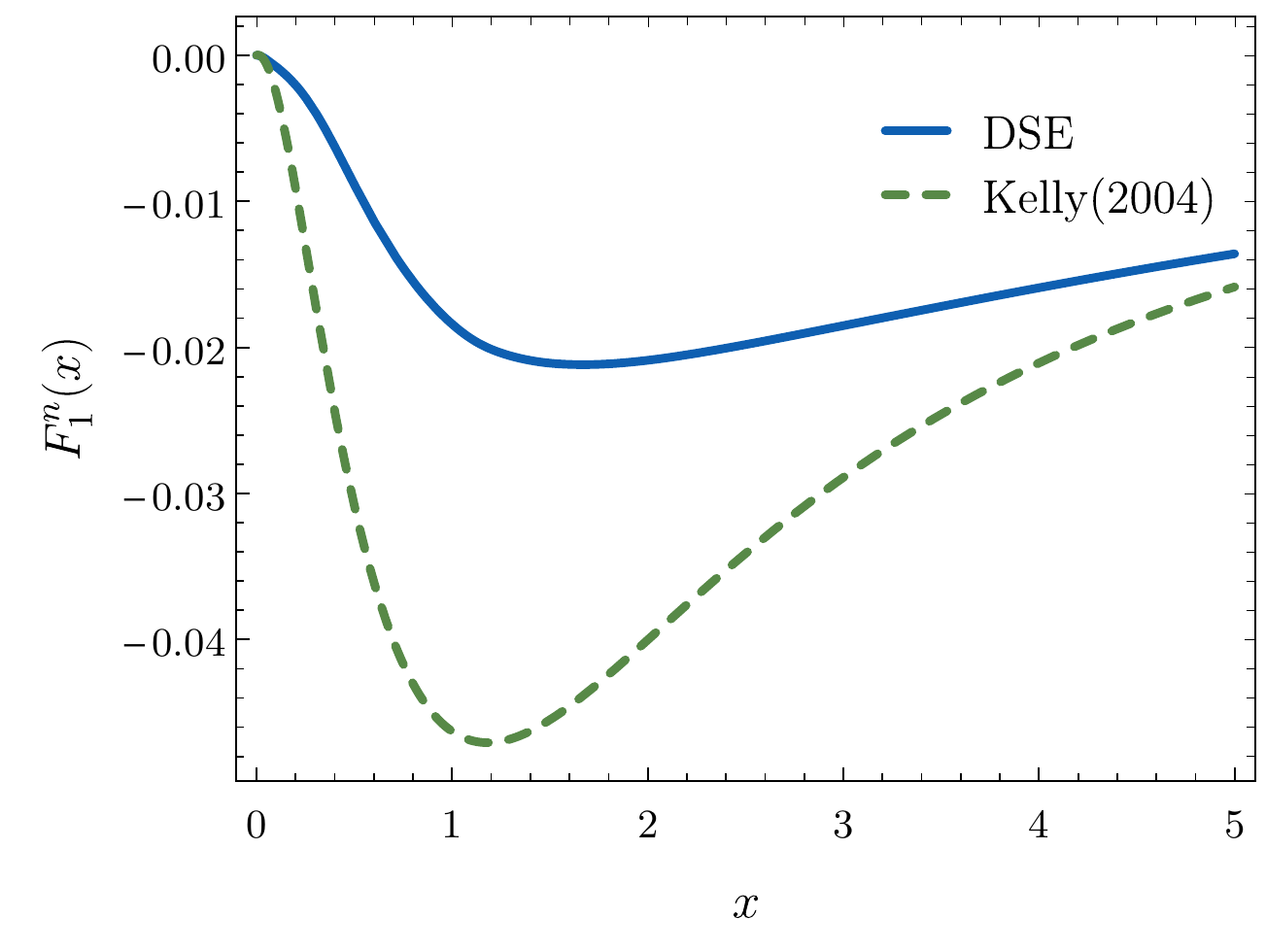}\hspace*{2ex } &
\includegraphics[clip,width=0.425\linewidth]{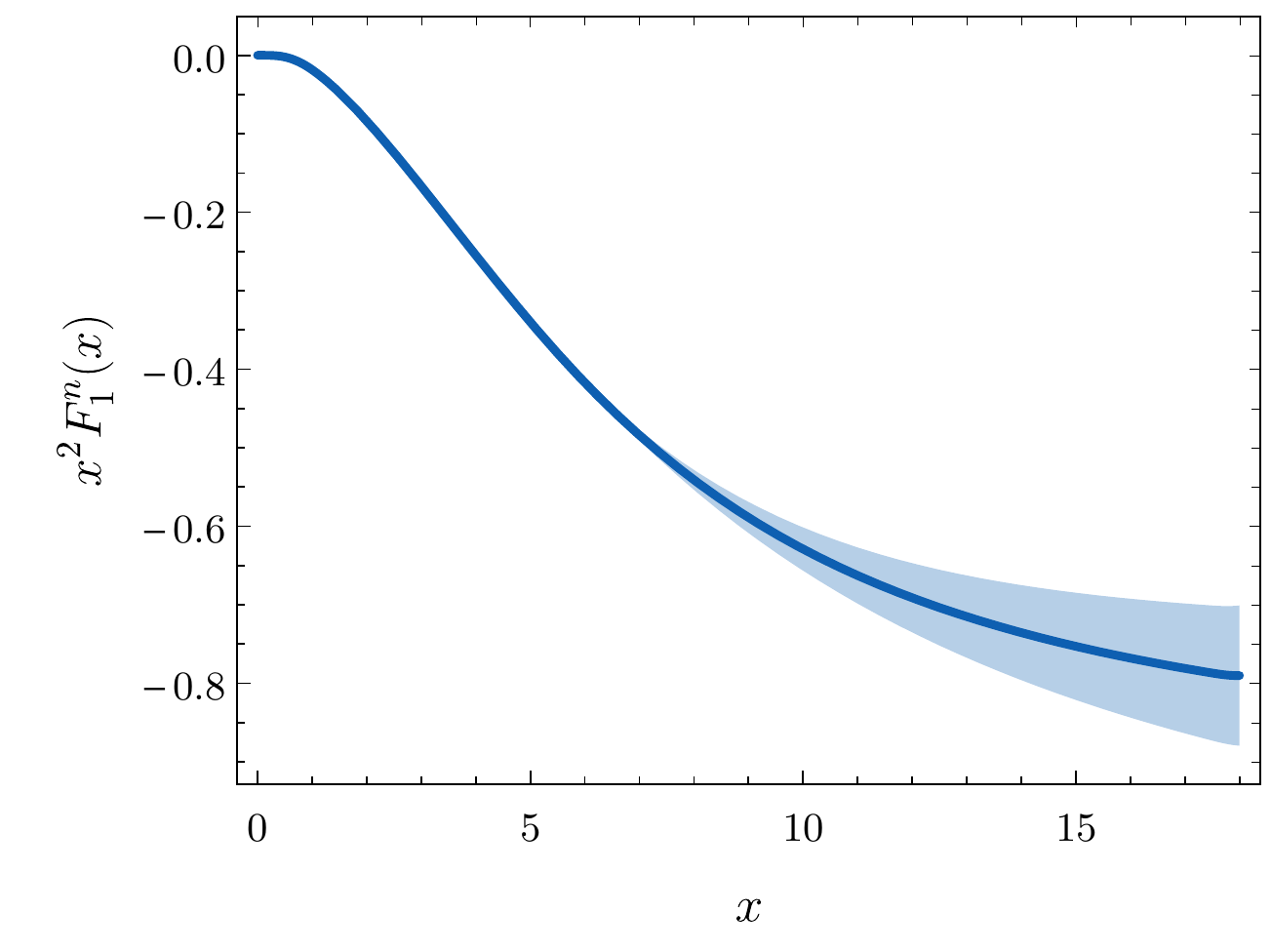}\vspace*{-0ex}
\end{tabular}
\begin{tabular}{lr}
\includegraphics[clip,width=0.425\linewidth]{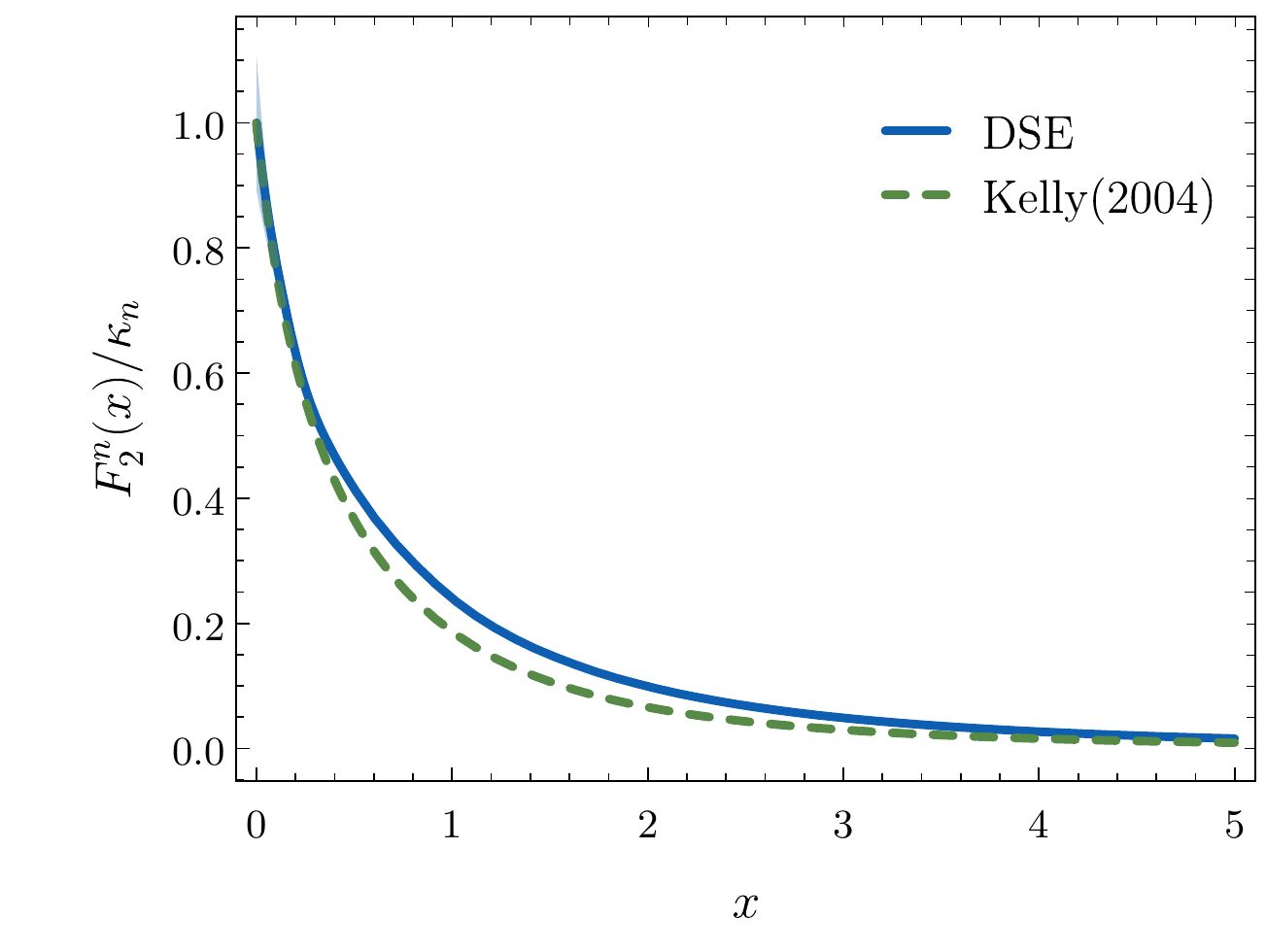}\hspace*{2ex } &
\includegraphics[clip,width=0.425\linewidth]{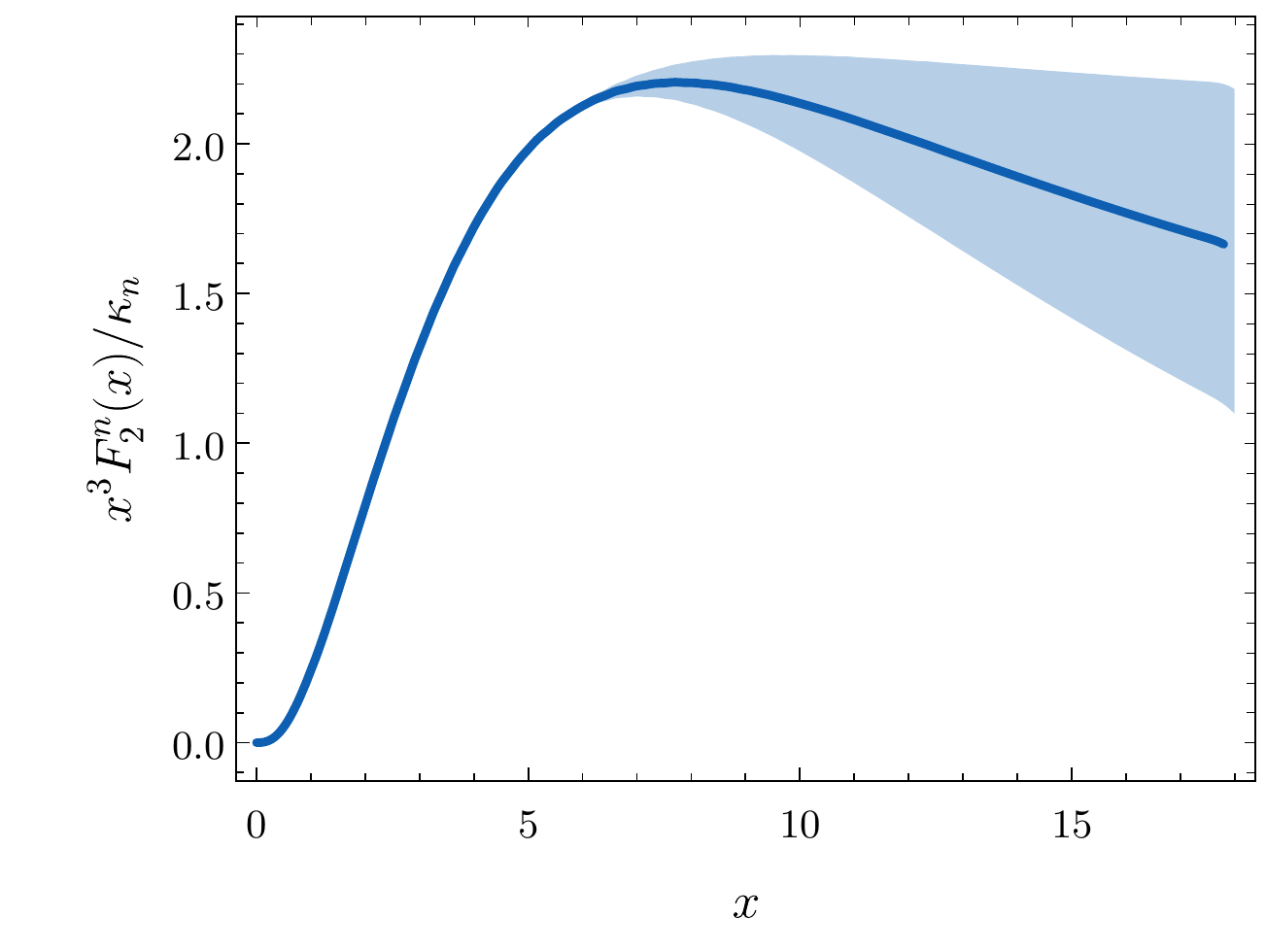}\vspace*{-1ex}
\end{tabular}
\end{center}
%
\caption{\label{FFDPn}
Neutron: herein, solid blue curves.
\emph{Upper panels} --  Dirac form factor, $F_1^n(x)$, $x=Q^2/m_N^2$;
and \emph{lower panels} -- Pauli form factor, $F_2^n(x)$.   ($\kappa_p = -1.59$.)
In both cases, the right panels depict $x^s$-weighted form factors, with $s$ chosen to match the associated na\"ive scaling dimension.
The empirical fits from Ref.\,\cite{Kelly:2004hm} are displayed in the left panels (dashed green curves).
In all panels, the $1\sigma$ band for the SPM approximants is shaded in light-blue.
}
\end{figure*}

\section{Elastic Form Factors}
\label{SecElastic}
\subsection{Dirac and Pauli}
Our result for the proton's Dirac form factor is depicted in Fig.\,\ref{FFDPp} -- upper panels.  The upper-left panel focuses primarily on the domain $x\in[0,x_{\rm max}]$, which was used to construct the SPM approximants.  The upper-right panel displays the approximants on $x\in[0,18]$.  The approximants satisfy our validation requirement, reproducing the direct calculation on $x\in[x_{\rm max} , 2x_{\rm max}]$.  On $x\in [2 x_{\rm max}, 4 x_{\rm max}=18]$, the result is a projection, obtained by extrapolation of the SPM approximants.  Our confidence in the extrapolation is indicated by the $1\sigma$ band: 68\% of all randomly generated SPM approximants produce a curve that lies within the blue shaded region.

\begin{table}[b]
\caption{\label{NStatic}
Faddeev equation results for a collection of nucleon static properties (Th).
Experimental (Ex) results taken from Ref.\,\cite{Tanabashi:2018oca, Antognini:1900ns, Pohl:2010zza}.  Dimensioned quantities in fm$^2$.
}
\begin{tabular}{lccccc}
Th & $\kappa_N$ & $r_E^2$ & $r_M^2$ & $r_E^2 M_N^2$ & $r_M^2 M_N^2$ \\\hline
$p$ & $\phantom{-}1.50$ & $\phantom{-}(0.76)^2$ & $(0.68)^2$ & $\phantom{-}(4.56)^2$ & $(4.08)^2$\\
$n$ & $-1.59$ & $-(0.26)^2$ & $(0.71)^2$ & $-(1.54)^2$ & $(4.24)^2$ \\\hline
Ex & $\kappa_N$ & $r_E^2$ & $r_M^2$ & $r_E^2 M_N^2$ & $r_E^2 M_N^2$ \\\hline
$p$ & $\phantom{-}1.79$ & $\phantom{-}(0.84)^2$ & $(0.84)^2$ & $\phantom{-}(4.00)^2$ & $(4.00)^2$ \\
$n$ & $-1.91$ & $-(0.34)^2$ & $(0.89)^2$ & $-(1.62)^2$ & $(4.24)^2$\\\hline
\end{tabular}
\end{table}

Context for our $F_1^p(x)$ result is provided by the dashed green curve in Fig.\,\ref{FFDPp} -- upper-left panel, which is the parametrisation of experimental data presented in Ref.\,\cite{Kelly:2004hm}.  Evidently, there is semi-quantitative agreement.  The slight mismatch at lower momenta can be attributed to the fact that both our Faddeev equation kernel and nucleon current do not contain resonant contributions, \emph{viz}.\ meson-cloud effects are omitted and the dressed-photon-quark vertex underestimates the contribution of the $\rho$-meson pole on $x\simeq 0$ \cite{Roberts:2000aa}.

The proton's Pauli form factor is plotted in Fig.\,\ref{FFDPp} -- lower panels, normalised by the proton's anomalous magnetic moment, listed in Table~\ref{NStatic}.  Once again, the approximants satisfy our validation requirement, reproducing the direct calculation on $x\in[x_{\rm max} , 2x_{\rm max}]$; and the result on $x\in [2 x_{\rm max}, 4 x_{\rm max}]$ is obtained by extrapolation of the SPM approximants.

The comparison with experimental data, represented by the parametrisations in Ref.\,\cite{Kelly:2004hm}, is slightly less favourable at lower momenta than for $F_1^p$.  This is because $F_2^p$ is a magnetic form factor, sensitive to angular momentum; hence, omission of meson cloud effects is more noticeable.  Naturally, cloud effects diminish rapidly with increasing $x$.  These same features are also apparent in the transition form factors nucleon-to-Roper \cite{Segovia:2015hra, Chen:2018nsg, Burkert:2019bhp} and nucleon-to-$\Delta$ \cite{Eichmann:2011aa, Segovia:2013rca, Lu:2019bjs}.

The right panels of Fig.\,\ref{FFDPp} reveal that whilst $F_1^p(x)$ exhibits incidental scaling-like behaviour on $x\gtrsim 5$, $F_2^p(x)$ does not.  This difference can lead to a zero in the proton's Sachs electric form factor.

\begin{figure*}[!t]
\begin{center}
\begin{tabular}{lr}
\includegraphics[clip,width=0.425\linewidth]{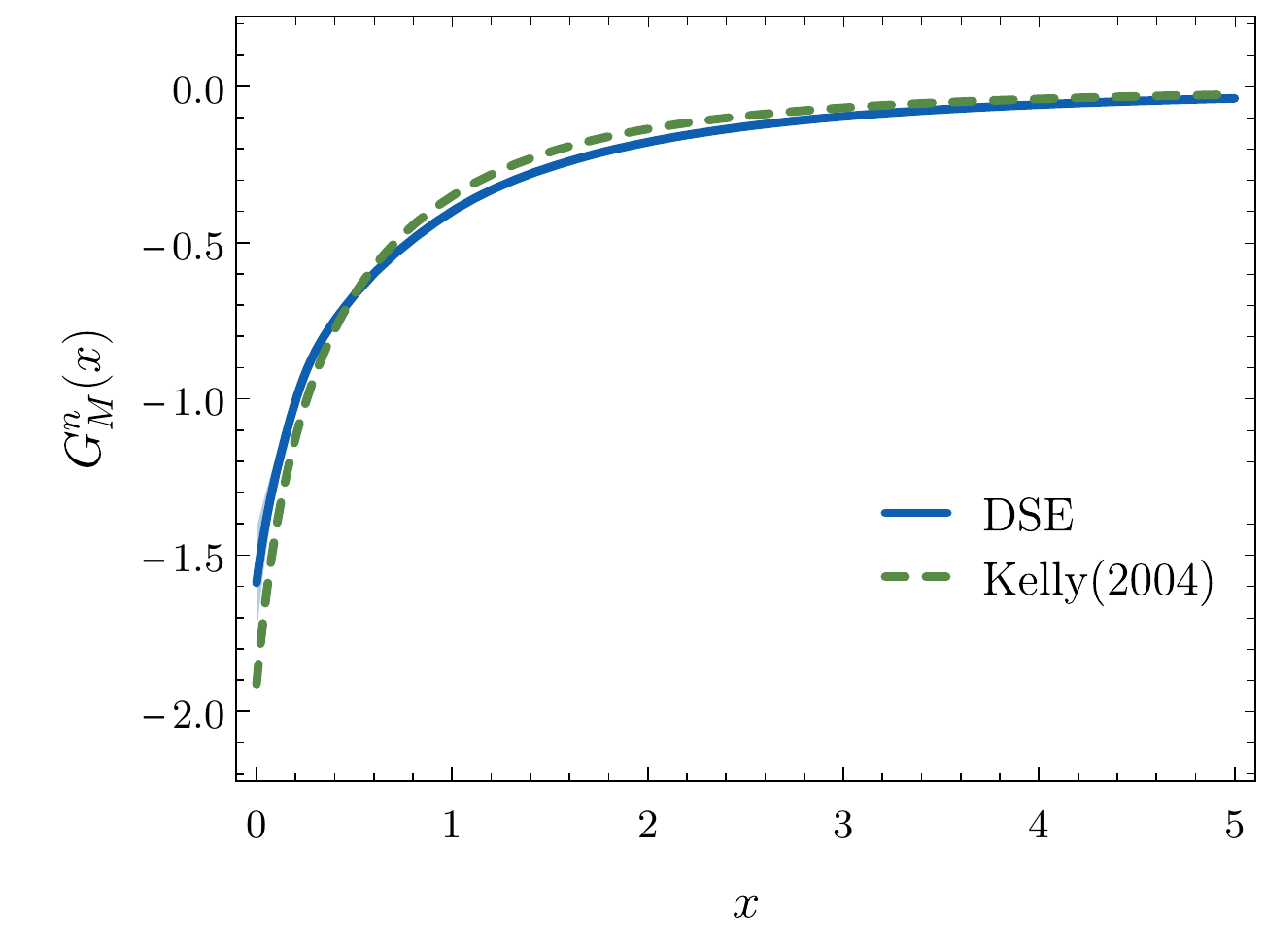}\hspace*{2ex } &
\includegraphics[clip,width=0.425\linewidth]{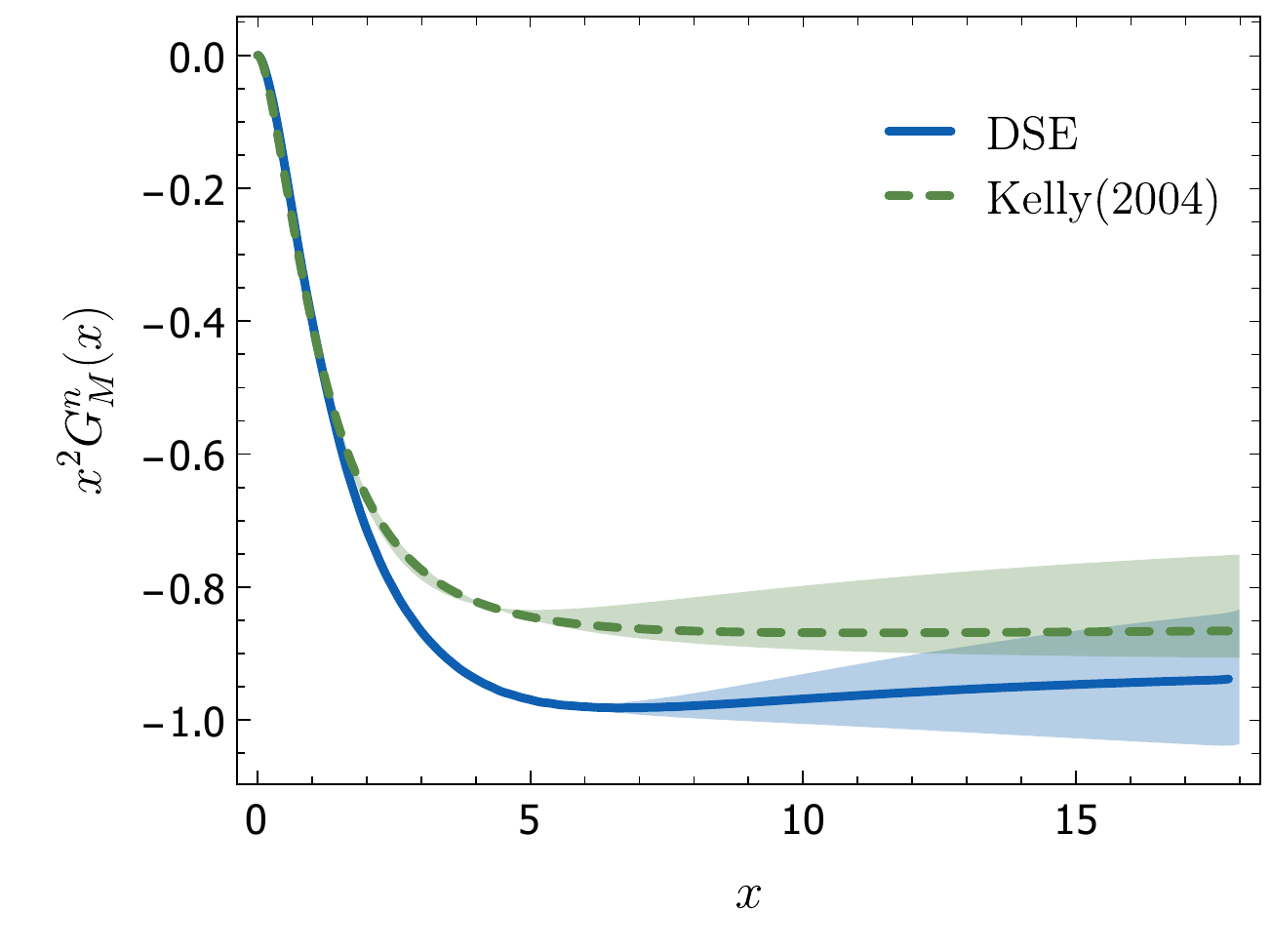}\vspace*{-0ex}
\end{tabular}
\begin{tabular}{lr}
\includegraphics[clip,width=0.425\linewidth]{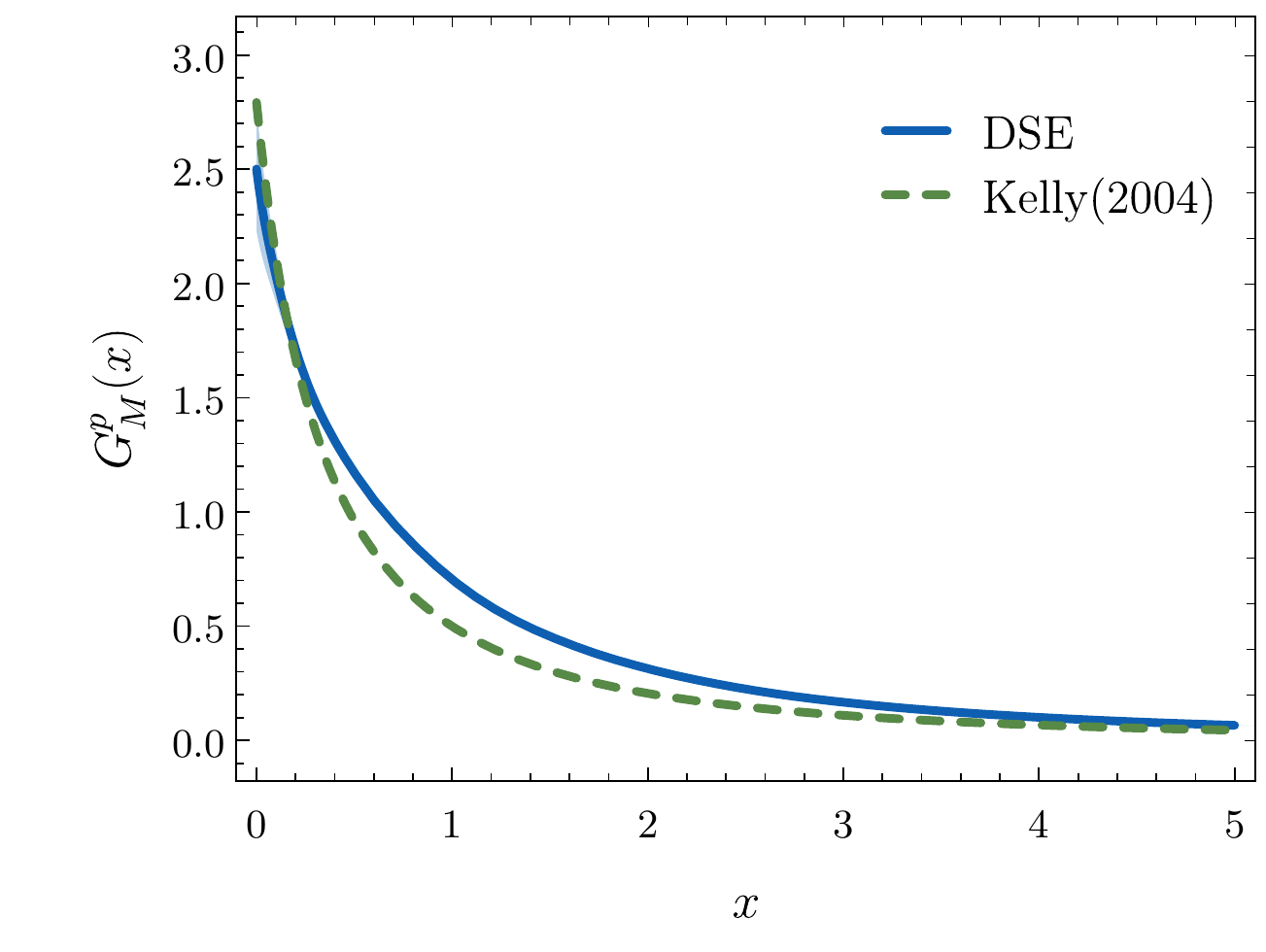}\hspace*{2ex } &
\includegraphics[clip,width=0.425\linewidth]{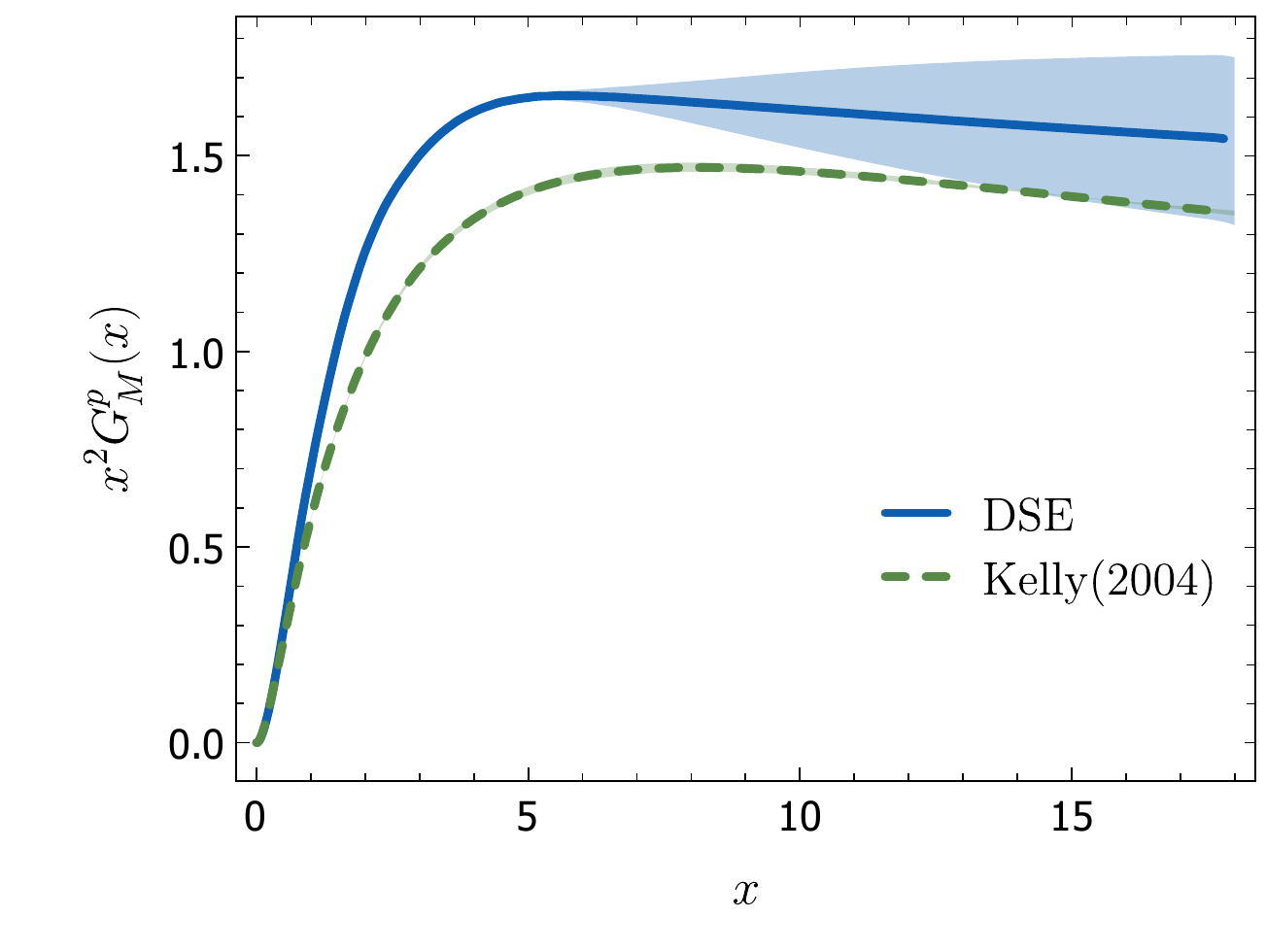}\vspace*{-1ex}
\end{tabular}
\end{center}
%
\caption{\label{GMnp}
Nucleon Sachs magnetic form factors.
\emph{Upper panels} --  Neutron. Left panel,  Sachs magnetic form factor, $G_M^n(x)$, $x=Q^2/m_N^2$; and right panel, $x^2 G_M^n(x)$.
\emph{Lower panels} -- Proton.  Left panel. Sachs magnetic form factor, $G_M^p(x)$;  and right panel, $x^2 G_M^p(x)$.
All panels.  The $1\sigma$ band for the SPM approximants is shaded in light-blue.
\emph{N.B}.\ Both right panels display $x^2 G_M(x)$; hence, agreement at the level of 0.1\% between theory and experiment is expressed in the image as an absolute difference of $0.1$ at $x=10$.
The fits to experimental data in Ref.\,\cite{Kelly:2004hm} are depicted as dashed green curves, with the quoted uncertainty on the fits drawn as shaded green bands in the right panels.
}
\end{figure*}

The neutron's elastic form factors are depicted in Fig.\,\ref{FFDPn}.  On the low-$x$ domain, the Dirac form factor agrees qualitatively with the parametrisation of experimental data.  There are differences in detail, again owing to omission of resonant contributions in our calculation.  They appear accentuated here because the neutron's Dirac form factor is the charge-weighted sum of much larger, positive quantities, with cancellations producing the uniformly small form factor.

It is worth recalling here that in the limit of SU$(4)$ spin-flavour symmetry, the wave functions for $u$- and $d$-quarks within the neutron are identical and the interaction current notices only the difference between the electric charges; hence, $F_1^n(x) \equiv 0$.  Our nonzero result for $F_1^n(x)$ is thus evidence of SU$(4)$ spin-flavour symmetry-breaking in our formulation.  This is a typical outcome of the Poincar\'e-covariant Faddeev equation treatment of the nucleon \cite{Eichmann:2009qa, Wang:2018kto, Qin:2019hgk}, which introduces correlations between the momentum-space behaviour of the solution and its spin-isospin structure; and a natural consequence of the presence of diquark correlations within the nucleon.

The right panels of Fig.\,\ref{FFDPn} highlight the large-$x$ behaviour of the neutron's form factors: there is no evidence for scaling in either case.  Consequently, the neutron's electric form factor may also exhibit a zero.

\begin{figure*}[!t]
\begin{center}
\begin{tabular}{lr}
\includegraphics[clip,width=0.425\linewidth]{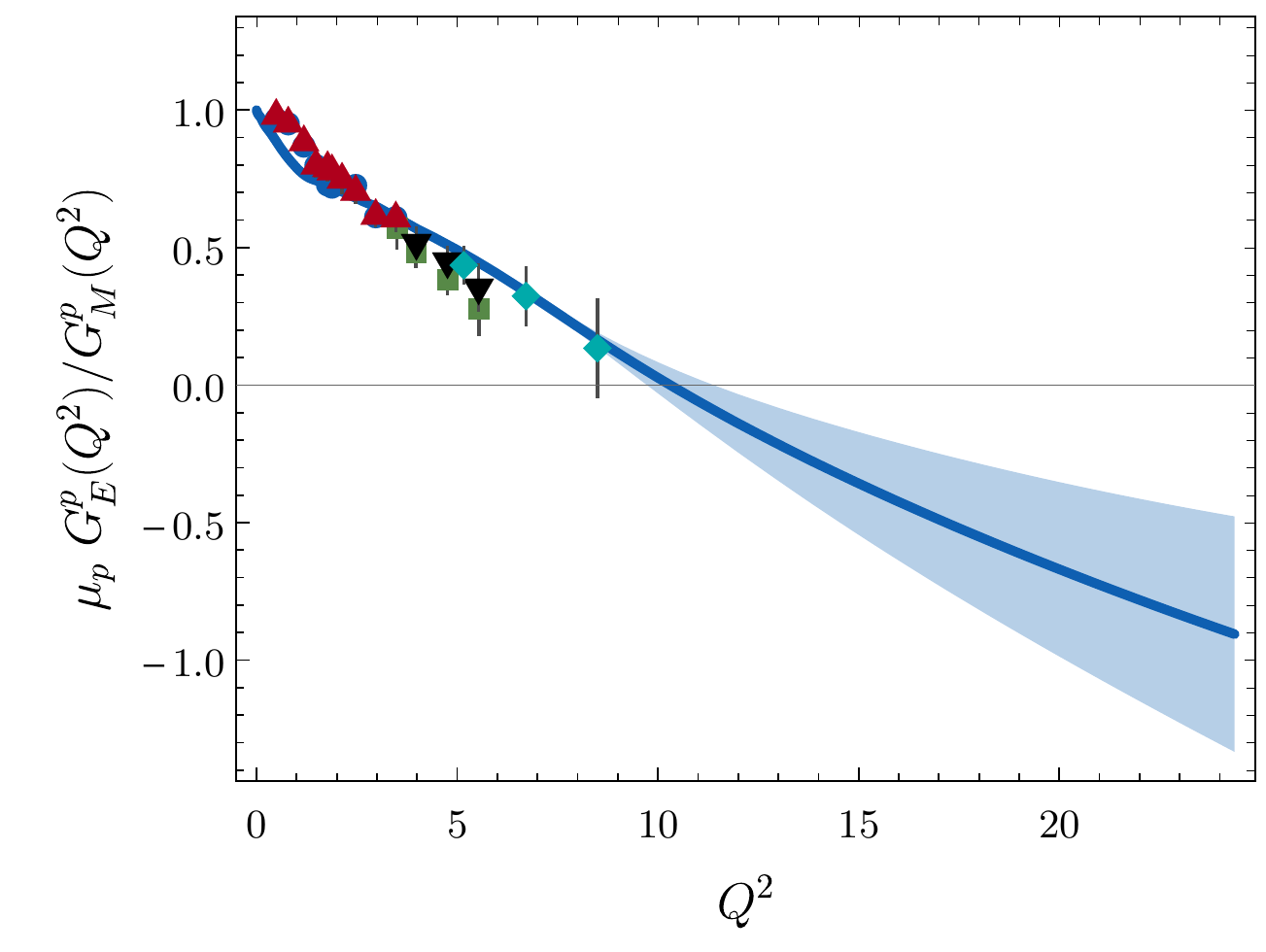}\hspace*{2ex } &
\includegraphics[clip,width=0.425\linewidth, height=0.24\textheight]{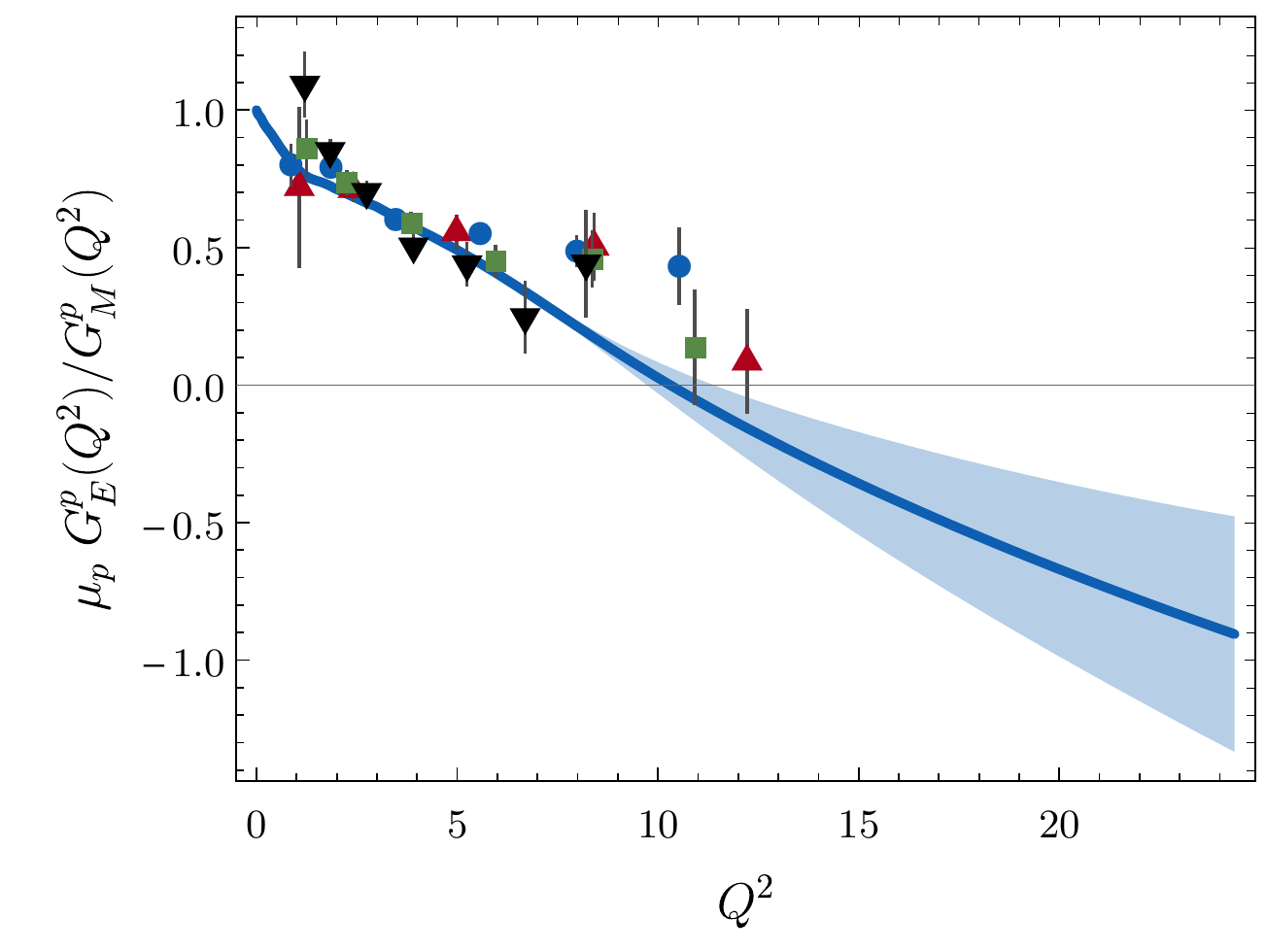}\vspace*{-0ex}
\end{tabular}
\begin{tabular}{lr}
\includegraphics[clip,width=0.425\linewidth]{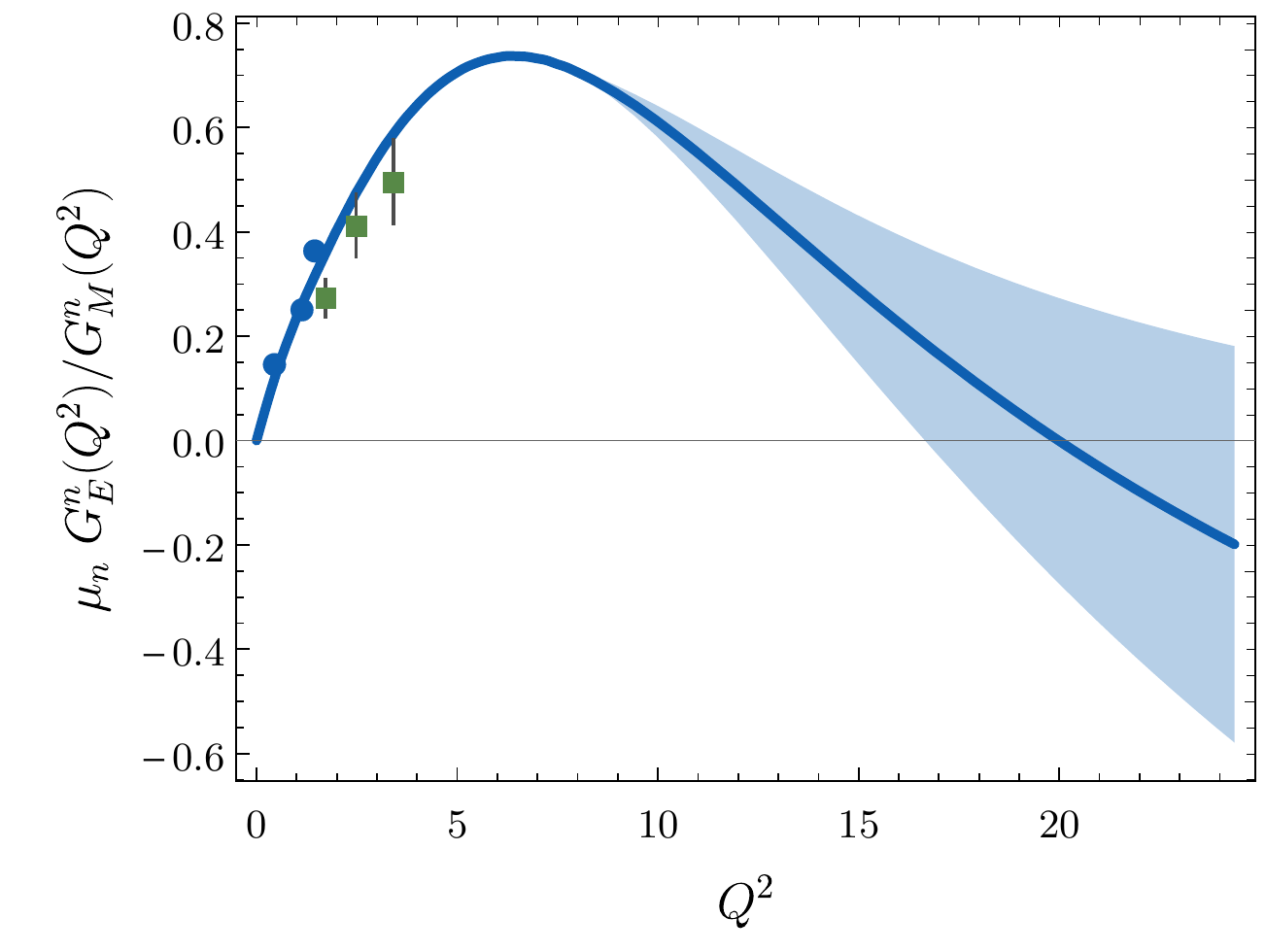}\hspace*{2ex } &
\includegraphics[clip,width=0.425\linewidth, height=0.24\textheight]{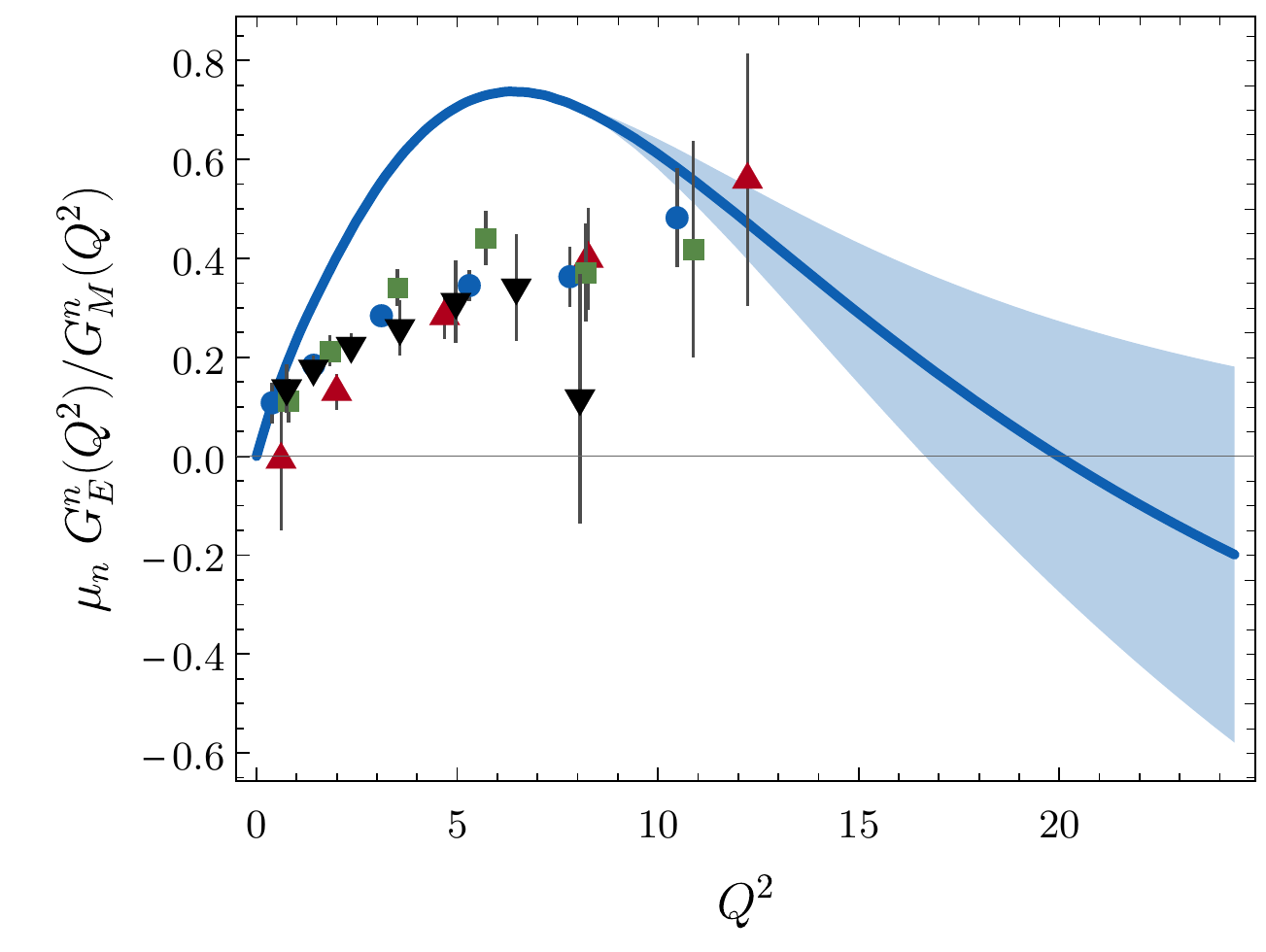}\vspace*{-1ex}
\end{tabular}
\end{center}
%
\caption{\label{GEonGM}
Ratios of Sachs form factors, $\mu_N G_E^N(x)/G_M^N(x)$.
\emph{Upper panels} --  Proton.  \underline{Left}, calculation herein compared with data
(red up-triangles \cite{Jones:1999rz};
green squares \cite{Gayou:2001qd};
blue circles \cite{Punjabi:2005wq};
black down-triangles \cite{Puckett:2010ac};
and cyan diamonds \cite{Puckett:2011xg});
\underline{right}, compared with available lQCD results, drawn from Ref.\,\cite{Kallidonis:2018cas}.
\emph{Lower panels} -- Neutron.  \underline{Left}, comparison with data (blue circles \cite{Madey:2003av} and green squares \cite{Riordan:2010id}); \underline{right}, with available lQCD results, drawn from Ref.\,\cite{Kallidonis:2018cas}.
In all panels, the $1\sigma$ band for the SPM approximants is shaded in light blue.
}
\end{figure*}

\subsection{Sachs Magnetic and Electric}
We depict the nucleon Sachs magnetic form factors in Fig.\,\ref{GMnp}: again, the results on $x\in [2 x_{\rm max}, 4 x_{\rm max}]$ are obtained by extrapolation of the SPM approximants.  In all cases, there is good agreement with the parametrisation of experimental data provided in Ref.\,\cite{Kelly:2004hm}.  This is especially true on $x\gtrsim 2$, \emph{i.e}.\ outside the domain on which resonant contributions to the Faddeev kernel and nucleon electromagnetic current are significant.  (Highlighting the sensitivity of form factors to the momentum dependence of the quark-quark interaction, such agreement is not achieved using a symmetry-preserving treatment of a momentum-independent/contact vector$\,\otimes\,$vector interaction \cite{Segovia:2014aza}.)  Regarding the fits, a paucity of data on $G_M^n$ at large-$x$ is evident in the uncertainty band on the green dashed curve displayed in Fig.\,\ref{GMnp}--upper right.

Data on $R_{EM}^p(Q^2)=\mu_p G_E^p(Q^2)/G_M^p(Q^2)$, obtained using polarisation transfer reactions at JLab, show a trend toward zero with increasing momentum-transfer-squared \cite{Jones:1999rz, Gayou:2001qd, Punjabi:2005wq, Puckett:2010ac, Puckett:2011xg}.  We depict this ratio and its analogue for the neutron in Fig.\,\ref{GEonGM}.

For the proton, our analysis predicts
\begin{equation}
\mu_p\, \frac{G_E^p(Q_z^2)}{G_M^p(Q_z^2)} = 0\,, Q_z^2 = 10.3^{+1.1}_{-0.7}\,{\rm GeV}^2\,.
\end{equation}
This value is compatible with, although a little larger than, that obtained earlier using the direct approach to calculating nucleon form factors \cite{Segovia:2014aza}: $Q_z^2 \approx 9.5\,$GeV$^2$; and a more recent inference based on $\rho$-meson elastic form factors \cite{Xu:2019ilh}: $Q_z^2 \approx 9.4(3)\,$GeV$^2$.

Regarding the neutron, Ref.\,\cite{Segovia:2014aza} predicted a peak in this ratio at $Q^2\approx 6\,$GeV$^2$, which is reproduced herein.  Furthermore, it located a zero at $Q_{z_n}^2\approx 12\,$GeV$^2$.  With our new method for reaching to large-$Q^2$, we find
\begin{equation}
 \mu_n\, \frac{G_E^n(Q_{z_n}^2)}{G_M^n(Q_{z_n}^2)} = 0\,, Q_{z_n}^2 = 20.1^{+10.6}_{-\phantom{1}3.5}\,{\rm GeV}^2\,,
\end{equation}
\emph{viz}.\ at $1\sigma$ confidence level, this ratio is likely to exhibit a zero, but it probably lies beyond the reach of 12\,GeV beams at JLab.  On the other hand, our prediction of a peak in $R_{EM}^n(Q^2)$, which is a harbinger of the zero in this ratio, can be tested at the 12\,GeV JLab.

It is worth providing a qualitative explanation for the behaviour of $R_{EM}^n(Q^2)=\mu_n G_E^n(Q^2)/G_M^n(Q^2)$.
Considering Eq.\,\eqref{GEGM} and given that $F_1^n(Q^2=0)=0$, then $R_{EM}^n$ must be zero at $Q^2=0$.
It increases with increasing $Q^2$ because $(-F_2^n(0)/[4 m_N^2])$ is large and positive, but the derivative of $F_1^n(Q^2)$ at $Q^2=0$ is small.
Studying Fig.\,\ref{FFDPn}, a peak can appear in $R_{EM}^n(Q^2)$ because $Q^2F_2^n(Q^2)$ itself has a peak whereas, on the displayed domain, $F_1^n(Q^2)$ is negative and increases steadily in magnitude.  For the same reason, $R_{EM}^n(Q^2)$ can possess a zero.

\begin{figure*}[!t]
\begin{center}
\begin{tabular}{lr}
\includegraphics[clip,width=0.425\linewidth]{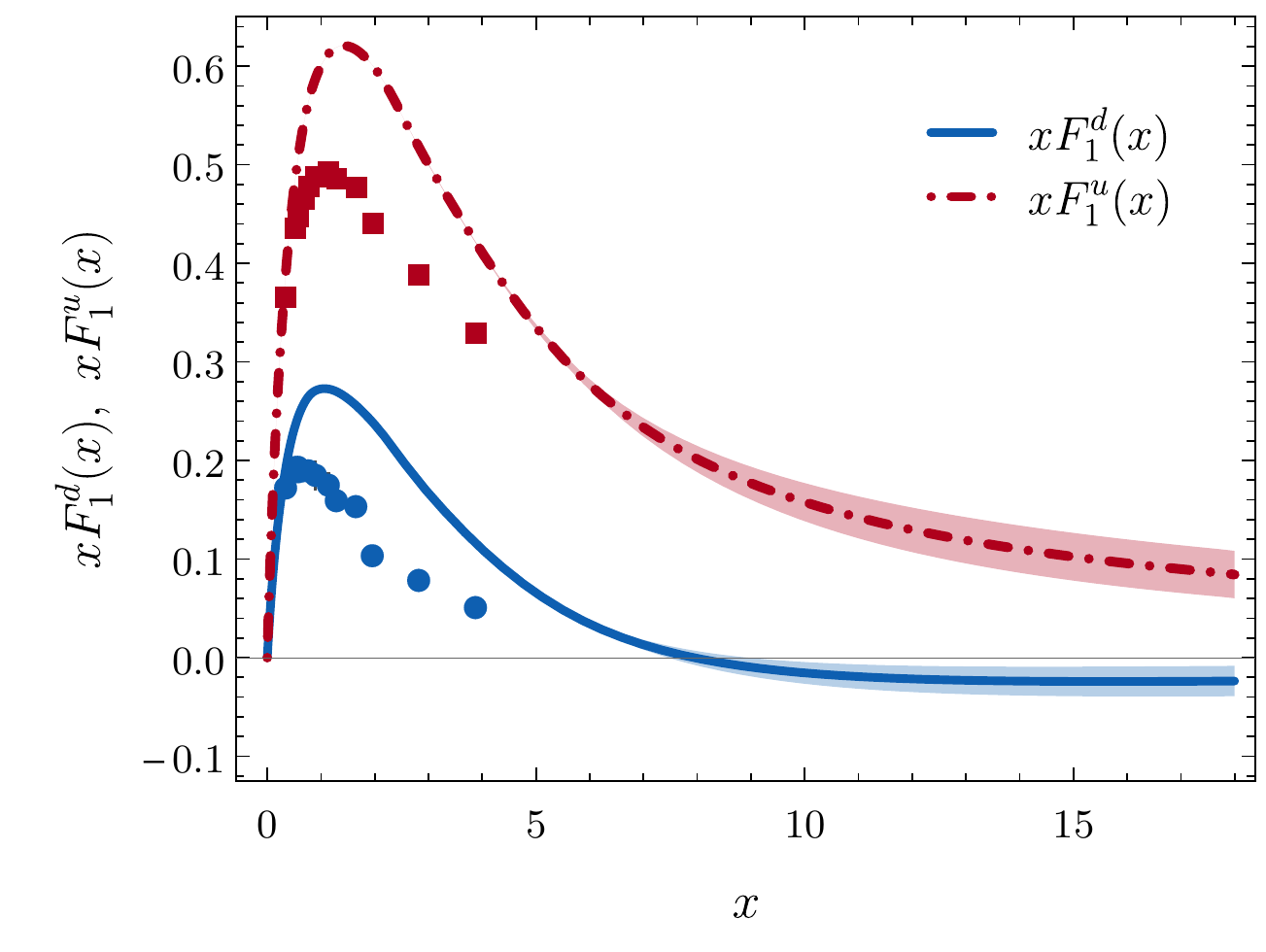}\hspace*{2ex } &
\includegraphics[clip,width=0.425\linewidth]{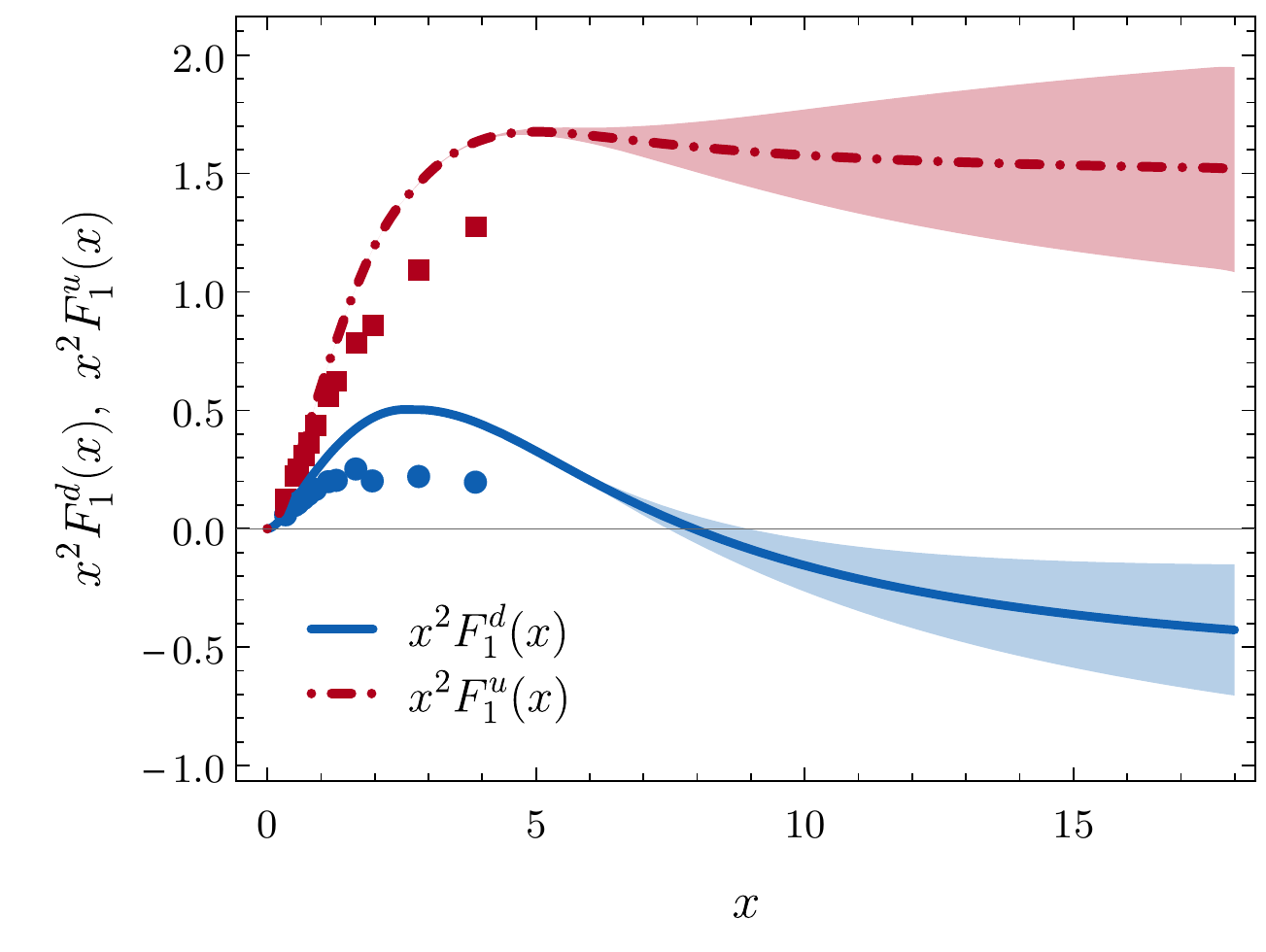}\vspace*{-0ex}
\end{tabular}
\begin{tabular}{lr}
\includegraphics[clip,width=0.425\linewidth]{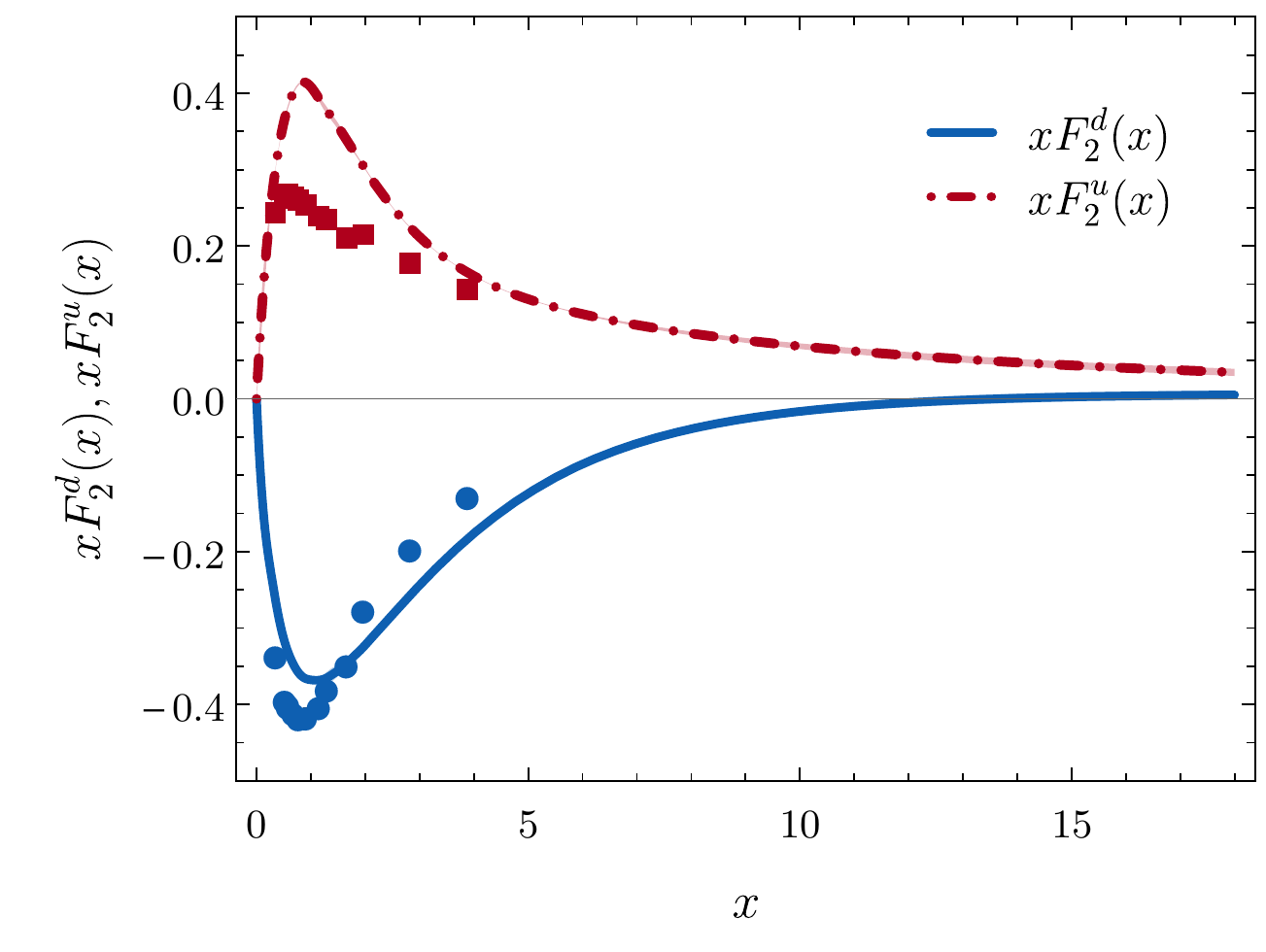}\hspace*{2ex } &
\includegraphics[clip,width=0.425\linewidth]{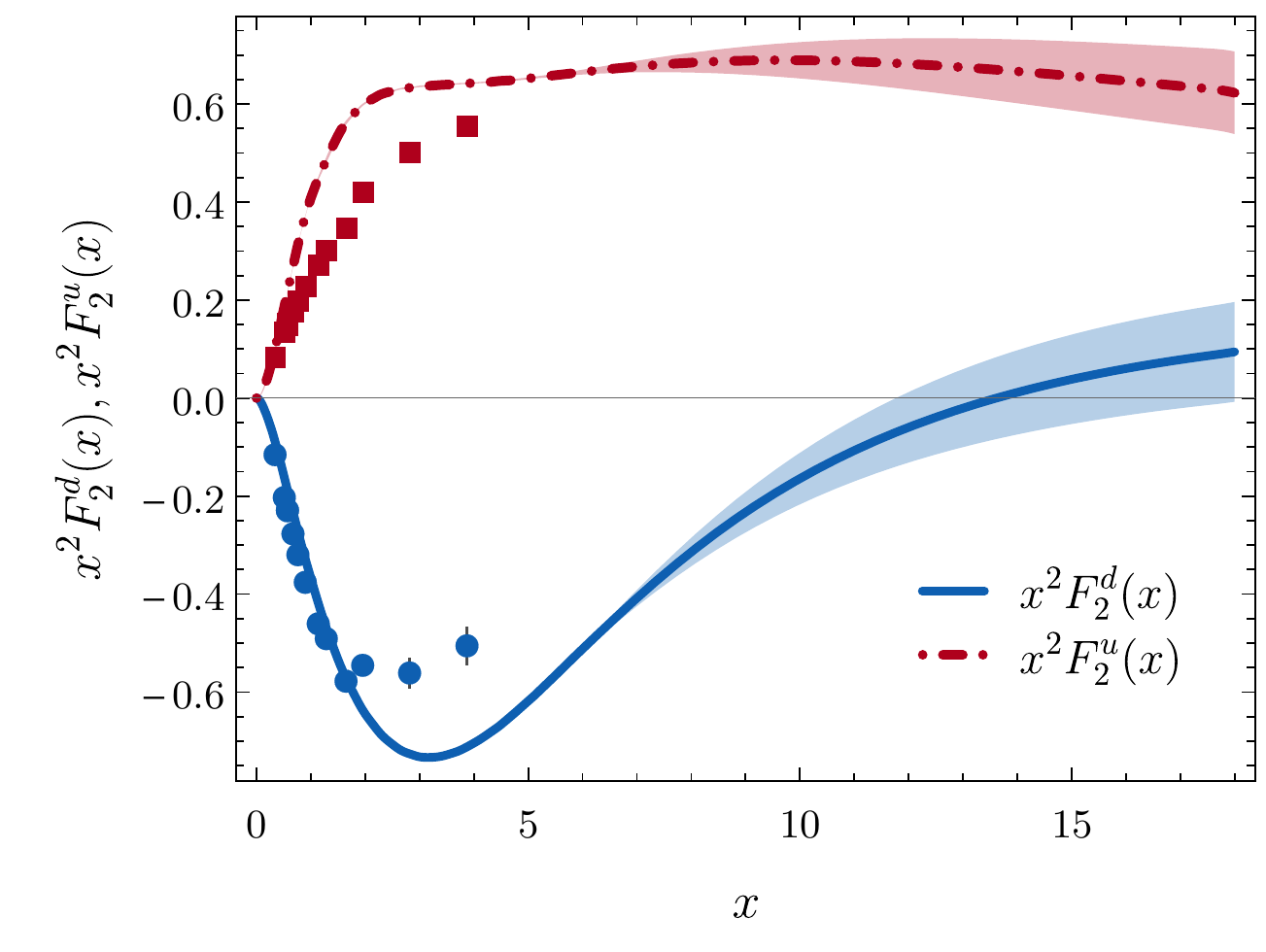}\vspace*{-1ex}
\end{tabular}
\end{center}
%
\caption{\label{FlavourSeparated}
Flavour separation.
\emph{Upper panels} -- $x$- and $x^2$-weighted Dirac form factors: $d$-quark, solid blue curve; and $u$-quark, dot-dashed red curve.
\emph{Lower panel} -- $x$- and $x^2$-weighted Pauli form factors: $d$-quark, solid blue curve; and $u$-quark, dot-dashed red curve.
In all panels -- shaded bands centred on each curve mark the $1\sigma$ uncertainty for the SPM approximants; and data from Ref.\,\cite{Cates:2011pz}: $d$-quark, blue circles; and $u$-quark, blue squares.
}
\end{figure*}

These remarks emphasise the importance of nucleon Pauli form factors, $F_2^{N=n,p}$, in connection with the appearance of a zero in $R_{EM}^N(Q^2)$.  As shown in Ref.\,\cite[Fig.\,4]{Segovia:2015ufa}, these form factors are very sensitive to those scalar-diquark components of the nucleon's rest-frame Faddeev wave function that carry nonzero quark+diquark orbital angular momentum.  Consequently, the appearance and location of a zero in $R_{EM}^N(Q^2)$ measures the strength of both quark-quark and angular momentum correlations within the nucleon.  Both are expressions of EHM.

\section{Flavour-separated Nucleon Form Factors}
\label{SecFlavour}
\subsection{Momentum Dependence}
Precision measurements of $G_E^n$ to $Q^2 =3.4\,$GeV$^2$ \cite{Riordan:2010id} enabled a flavour separation of the nucleon form factors on a sizeable momentum domain \cite{Cates:2011pz}.  Assuming one can neglect $s$-quark contributions:
\begin{equation}
\label{FlavourSep}
F_{i}^u = 2 F_{i}^p + F_{i}^{n}, \;
F_{i}^d = F_{i}^{p} + 2 F_{i}^{n} \,, i=1,2,
\end{equation}
where the Dirac and Pauli form factors are obtained from data using Eqs.\,\eqref{GEGM}.  Evidently,
\begin{equation}
\label{FlavourNorm}
F_{1}^{u}(x=0)=2\,, \; F_{1}^{d}(x=0)=1\,.
\end{equation}

Our results for the flavour separated form factors are presented in Fig.\,\ref{FlavourSeparated} along with the data from Ref.\,\cite{Cates:2011pz}.  In all cases, the calculations reproduce the trends seen in available data and agree semi-quantitatively in magnitudes.  Importantly, we predict that $F_{1,2}^d$ both exhibit a zero:
\begin{subequations}
\begin{align}
\label{F1dzero}
F_1^d(x_{z_1}) & = 0\,, x_{z_1} \approx \phantom{1}7.9_{-0.5}^{+1.2}\,,\\
F_2^d(x_{z_2}) & = 0\,, x_{z_2} \approx 13.6_{-1.9}^{+4.4}\,.
\end{align}
\end{subequations}
Although the zero in $F_1^d$ was seen previously \cite{Segovia:2014aza, Segovia:2015ufa}, the precision of these earlier calculations at large-$x$ was insufficient to reveal the zero in $F_2^d$.
New generation experiments at JLab are capable of testing Eq.\,\eqref{F1dzero}; indeed, results might soon be available \cite{Wojtsekhowski:2020tlo}.

Detailed explanations for the behaviour of the curves in Fig.\,\ref{FlavourSeparated} are presented elsewhere \cite{Segovia:2015ufa}.  Notwithstanding that, we recapitulate some aspects herein.

The upper panels of Fig.\,\ref{FlavourSeparated} show that $F_1^d$ is smaller than $F_1^u$, even allowing for the difference in normalisation, and decreases more quickly as $x$ increases.
This behaviour can be understood by noting that the proton's Faddeev wave function is dominated by the $[ud]$ scalar diquark correlation, which produces roughly $65$\% of the proton's normalisation \cite{Mezrag:2017znp}.  Hence, $ep$ scattering is dominated by the virtual photon striking the $u$-quark that is not participating in the correlation.  Scattering from the valence $d$-quark is suppressed because this quark is typically absorbed into a soft correlation.  The effect is particularly noticeable at large $x$.

As highlighted above, however, the proton also contains an active and measurable pseudovector diquark component.  This piece occurs in two combinations: $u\{ud\}$ and $d\{uu\}$.  The latter ensures that, although with smaller net probability, valence $d$-quarks are always available to participate in a hard scattering event; hence, it provides the leading contribution to $F_1^d$ on $x\gtrsim 2$.  As $x$ increases deeper into this domain, a zero is exposed in $F_1^d$.  Its location depends on interferences between the various diquark-component contributions to $ed$ scattering within the proton.  Thus, like the ratios of valence-quark parton distribution functions at large Bjorken-$x$ \cite{Roberts:2013mja, Segarra:2019gbp}, the location of the zero in $F_1^d$ is a measure of the relative probability of finding pseudovector and scalar diquarks in the proton.  Furthermore, and importantly, the existence of this zero highlights that any appearance of scaling in nucleon electromagnetic form factors on $x\lesssim 20$ is incidental because the zero expresses a continuing role for correlations that distinguish between quark flavours and impose different features upon their scattering patterns.

The lower panel of Fig.\,\ref{FlavourSeparated} depicts the proton's flavour-separated Pauli form factors.  The features already exposed by the behaviour of $F_1^{u,d}$ are also largely repeated here; and their explanations are similar.  There is one notable difference; namely, despite the fact that the $d$:$u$ ration in the proton is $2$:$1$, the $u$- and $d$-quark Pauli form factors are roughly equal in magnitude on $x\lesssim 5$.  We return to this in Sec.\,\ref{AnomMag}.

\begin{figure*}[!t]
\begin{center}
\begin{tabular}{lr}
\includegraphics[clip,width=0.425\linewidth]{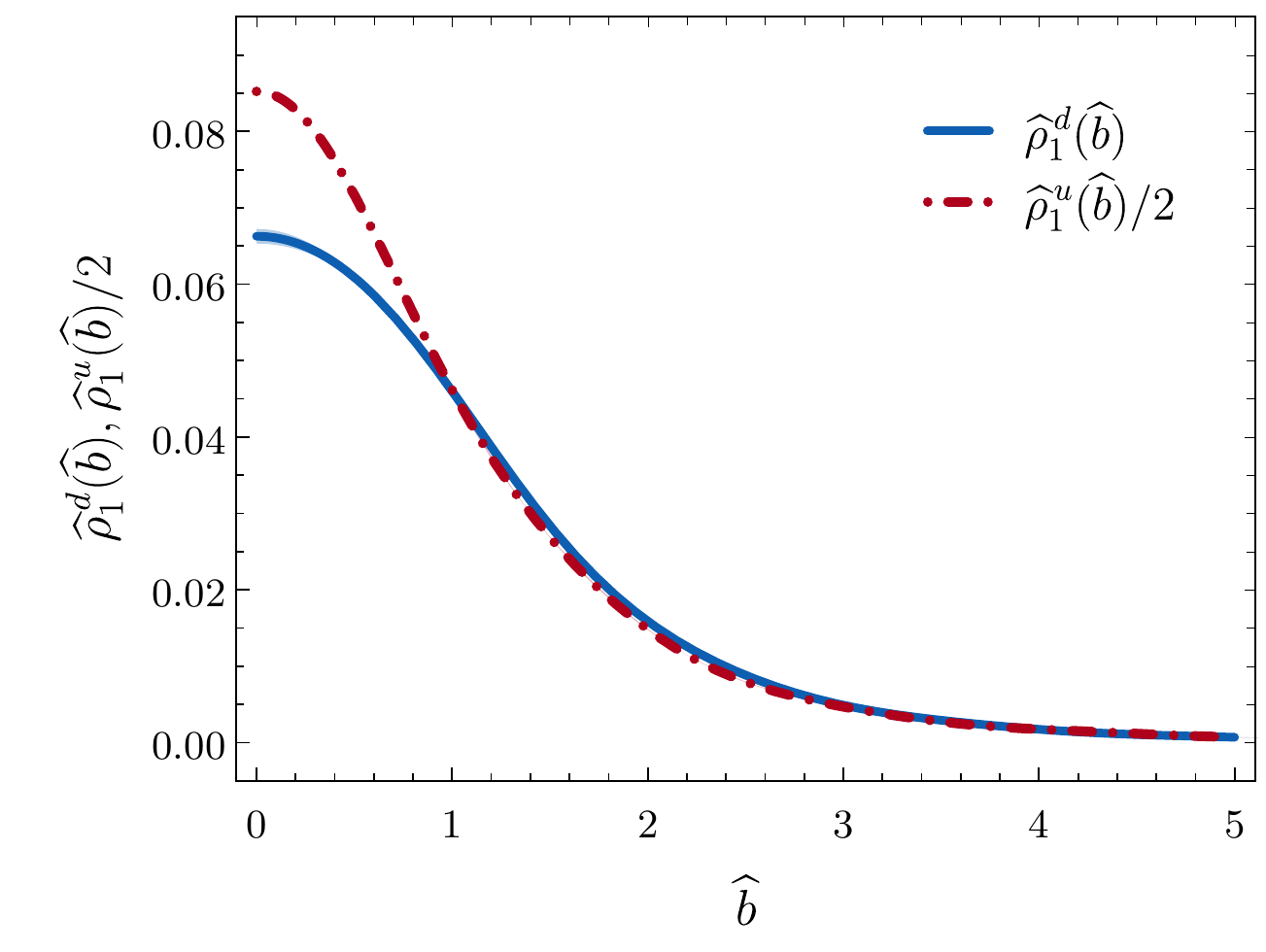}\hspace*{2ex } &
\includegraphics[clip,width=0.425\linewidth]{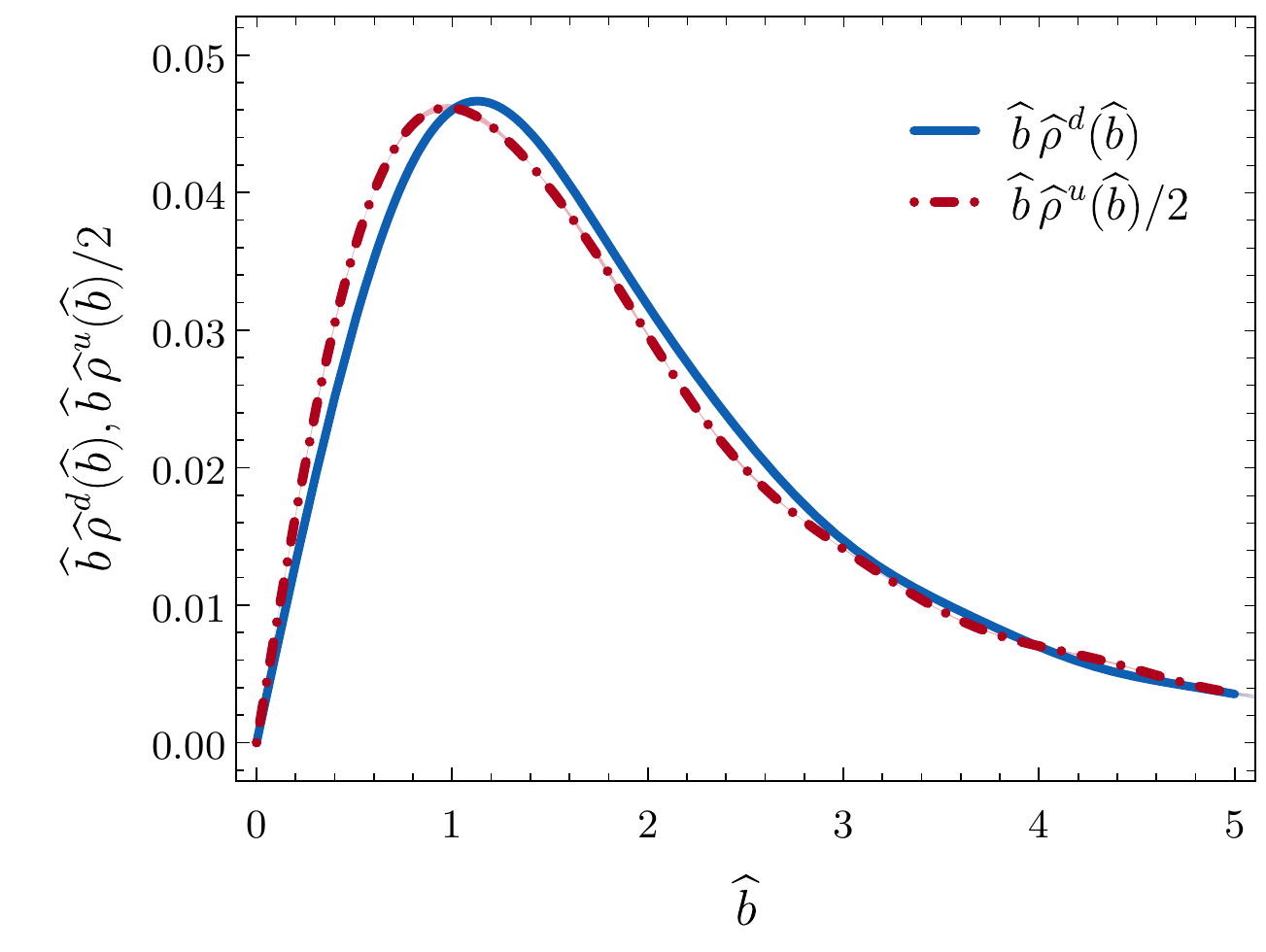}\vspace*{-0ex}
\end{tabular}
\begin{tabular}{lr}
\includegraphics[clip,width=0.425\linewidth]{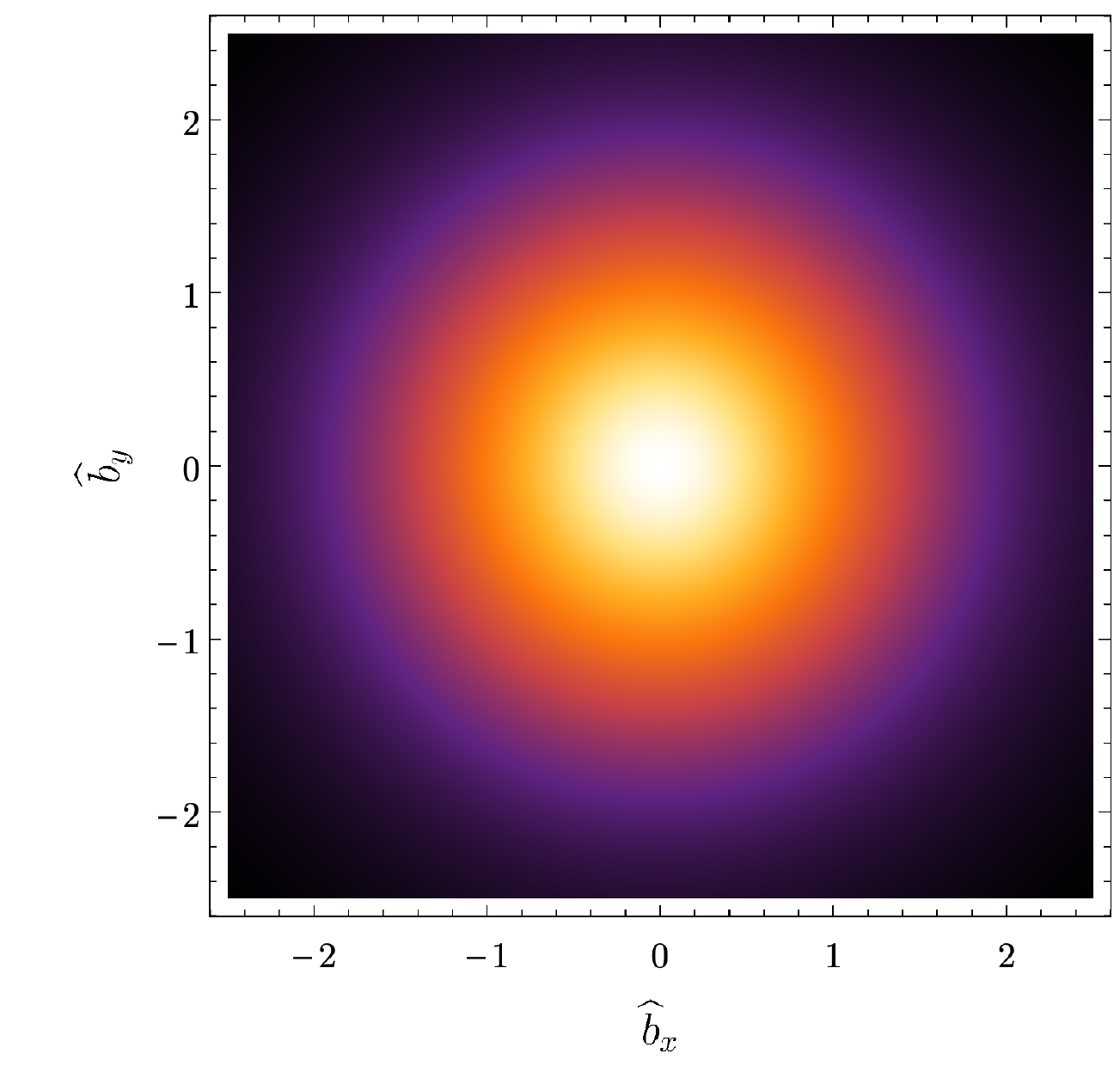}\hspace*{2ex } &
\includegraphics[clip,width=0.425\linewidth]{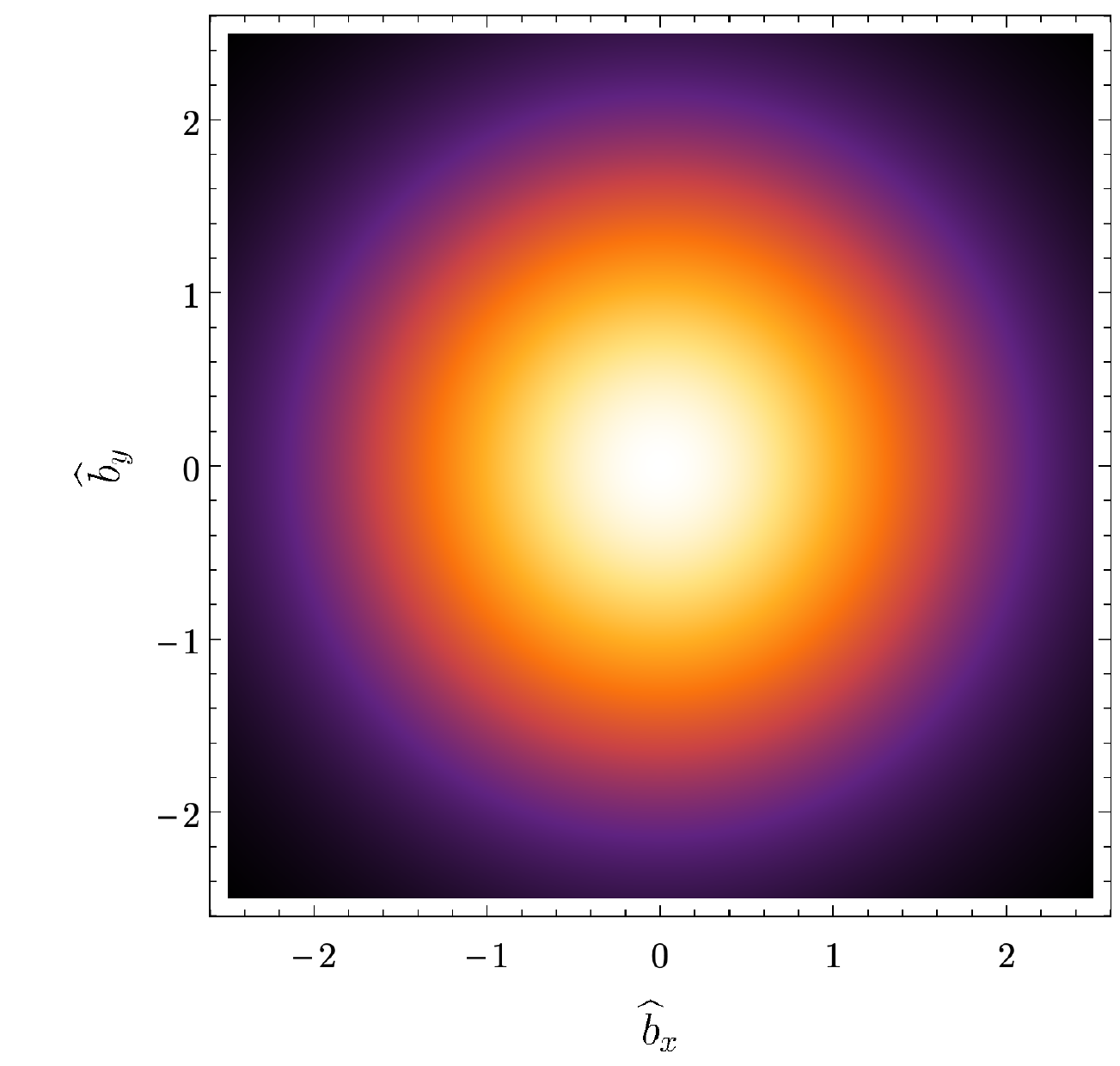}\vspace*{-1ex}
\end{tabular}
\end{center}
%
\caption{\label{Figfdensity}
Light-front transverse valence $u$- and $d$-quark densities.
\underline{Upper panels} -- one-dimensional profiles defined in Eq.\,\eqref{fdensity}: \emph{left},  $\hat\rho_1^f(|\hat b|)/2$; and \emph{right}, $|\hat b|\hat\rho_1^f(|\hat b|)$.
\underline{Lower panels} -- two dimensional images derived from the curves in the upper-left panel: \emph{left}, $\hat\rho_1^u(|\hat b|)/2$; and \emph{right}, $\hat\rho_1^d(|\hat b|)$.
\emph{N.B}.\ In all panels, the $u$-quark density is halved, eliminating the overall $2u$:$1d$ feature of the proton, so that a direct comparison between the $|\hat b|$-dependence of the densities can readily be made.
}
\end{figure*}

\subsection{Light-front-transverse Densities}
\label{SubSecLFTDs}
\subsubsection{Number}
Considering Eq.\,\eqref{FlavourNorm}, one can develop a perspective from which the flavour-separated elastic form factors can be viewed as providing a measure of the $Q^2$-dependence of valence $u$- and $d$-quark elastic scattering probabilities within the proton.  Such a connection can be made rigorous using generalised parton distributions in impact-parameter space so that the following two-dimensional Fourier transforms express light-front-transverse valence-quark densities \cite{Burkardt:2002hr, Diehl:2003ny, Carlson:2007xd, Mezrag:2016hnp} $(f=u,d)$:
\begin{equation}
\label{fdensity}
\hat \rho_1^f(|\hat b|) = \int\frac{d^2 \vec{q}_\perp}{(2\pi)^2}\,{\rm e}^{i \vec{q}_\perp \cdot \hat b}
F_1^{f}(Q^2)\,,
\end{equation}
with $F_1^f(Q^2)$ interpreted in the frame defined by $Q^2 = m_N^2 |q_\perp|^2$, $m_N q_\perp =  (Q_1,Q_2,0,0)$.  Such densities have also been considered elsewhere \cite{Miller:2010nz, GonzalezHernandez:2012jv, Mondal:2016xpk}.

For the purpose of visualising the proton's internal structure, it is instructive to display the valence-quark densities defined by Eq.\,\eqref{fdensity}; and this is done in Fig.\,\ref{Figfdensity}.  We have used dimensionless quantities and one can map into physical units using:
\begin{equation}
\rho_1^f(|b| =| \hat b|/m_N) = m_N^2 \,  \hat \rho_1^f(|\hat b|)\,,
\end{equation}
\emph{viz}.\ $|\hat b| = 1$ corresponds to $|b| \approx 0.2\,$fm and $\hat\rho_1^f=0.1 \Rightarrow \rho_1^f\approx 2.3/{\rm fm}^2$.

The images in Fig.\,\ref{Figfdensity} provide a good amount of information, which we now detail.

(\emph{i}) Fig.\,\ref{Figfdensity}\,--\,upper-left-panel shows that valence $u$-quarks are more likely than valence $d$-quarks to be found near the proton's centre of transverse momentum (CoTM).  (We have divided $\hat\rho_1^u(\hat b)$ by two in order to eliminate the simple $d$:$u=1$:$2$ counting factor.)  The excess valence $u$-quark density lies on $|\hat b| \leq 1 \Rightarrow |b| \lesssim 0.2\,$fm, with $\hat\rho_1^d(0)/\hat\rho_1^u(0) = 0.38$.  Both the size of the excess and extent of the associated domain are connected with the relative strength of scalar and pseudovector diquark correlations within the proton: omitting the pseudovector diquark, the $u$-quark excess is significantly larger.  The excess is also correlated with the fact that the ratio of valence-quark parton distribution functions at Bjorken-$x \simeq 1$ satisfies \cite{Roberts:2013mja}:
\begin{equation}
0<d_v(x_{\rm Bj})/u_v(x_{\rm Bj})\approx 0.23 < 1/2\,.
\end{equation}

(\emph{ii}) Fig.\,\ref{Figfdensity}\,--\,upper-right-panel indicates that the $u$- and $d$-quark transverse densities have roughly the same mean radius.  This can be quantified by considering
\begin{equation}
\label{TransverseRadii}
(\hat \lambda_1^f)^2 := \int\! \frac{d^2\hat b}{(2\pi)^4}\, |\hat b|^2 \hat \rho_1^f(|\hat b|)
=\left.  \frac{ - 4}{F_1^f(0)} \frac{d F_1^f(x)}{dx} \right|_{x=0}\,,
\end{equation}
from which we find
\begin{equation}
\hat \lambda_1^u \approx 3.5 \approx \hat \lambda_1^d =: \hat \lambda_1,
\end{equation}
a value corresponding to roughly $0.7\,$fm.   (One has the same outcome empirically, with $\hat\lambda_1 \approx 3.3$ \cite{Kelly:2004hm}.)
Notwithstanding this, higher moments of the distributions reveal that the $d$-quark density exceeds the $u$-quark density on $1\lesssim |\hat b|\lesssim \hat \lambda_1^u$.  This is apparent in the figures and highlighted by the two-dimensional plots in the lower panels of Fig.\,\ref{Figfdensity}, which reveal that when compared with the $u$-quark density, the $d$-quark density is a little more diffuse.  This had to be the case, given that both curves in Fig.\,\ref{Figfdensity}\,--\,upper-left-panel integrate to unity and the $u$-quark curve has greater support on $|\hat b|\lesssim 1$.

(\emph{iii}) The upper panels of Fig.\,\ref{Figfdensity} show that the valence $u$- and $d$-quark densities have the same behaviour on $|\hat b|\gtrsim\hat \lambda_1 $, \emph{viz}.\ they have practically identical long-range tails.  This feature ensures that the proton's quark core appears as a rotationally-invariant unit-charge object to a long-wavelength photon probe and the neutron's core, as a rotationally invariant zero-charge target.

(\emph{iv}) Using Eq.\,\eqref{FlavourSep} and the results in Fig.\,\ref{Figfdensity}, one immediately perceives that the transverse density associated with the neutron's dressed-quark core is: negative on $|\hat b|\lesssim 1$; positive on $1\lesssim |\hat b|\lesssim 3.5$; and approximately zero thereafter.

\begin{figure*}[!t]
\begin{center}
\begin{tabular}{lr}
\includegraphics[clip,width=0.425\linewidth]{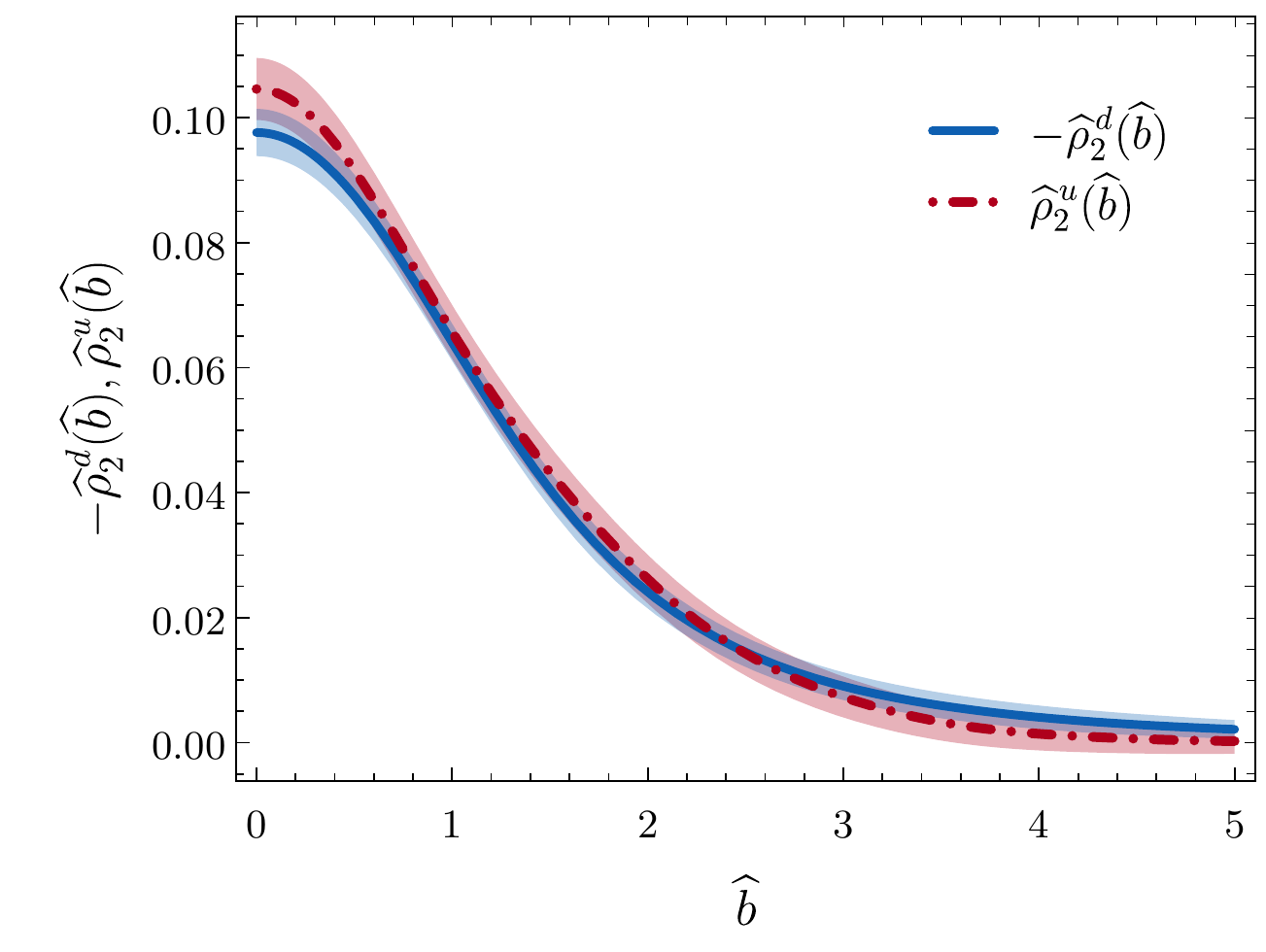}\hspace*{2ex } &
\includegraphics[clip,width=0.425\linewidth]{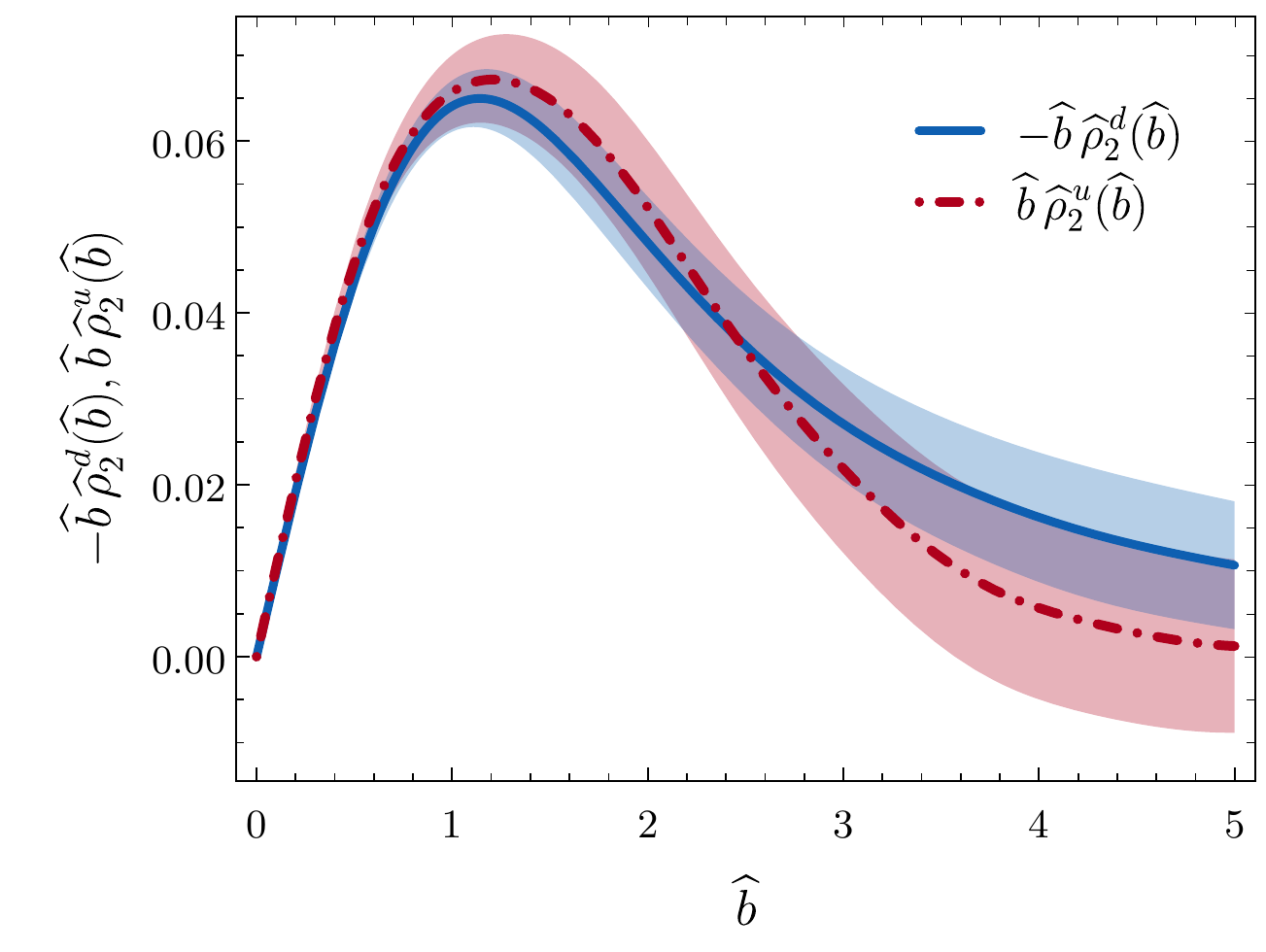}\vspace*{-0ex}
\end{tabular}
\begin{tabular}{lr}
\includegraphics[clip,width=0.425\linewidth]{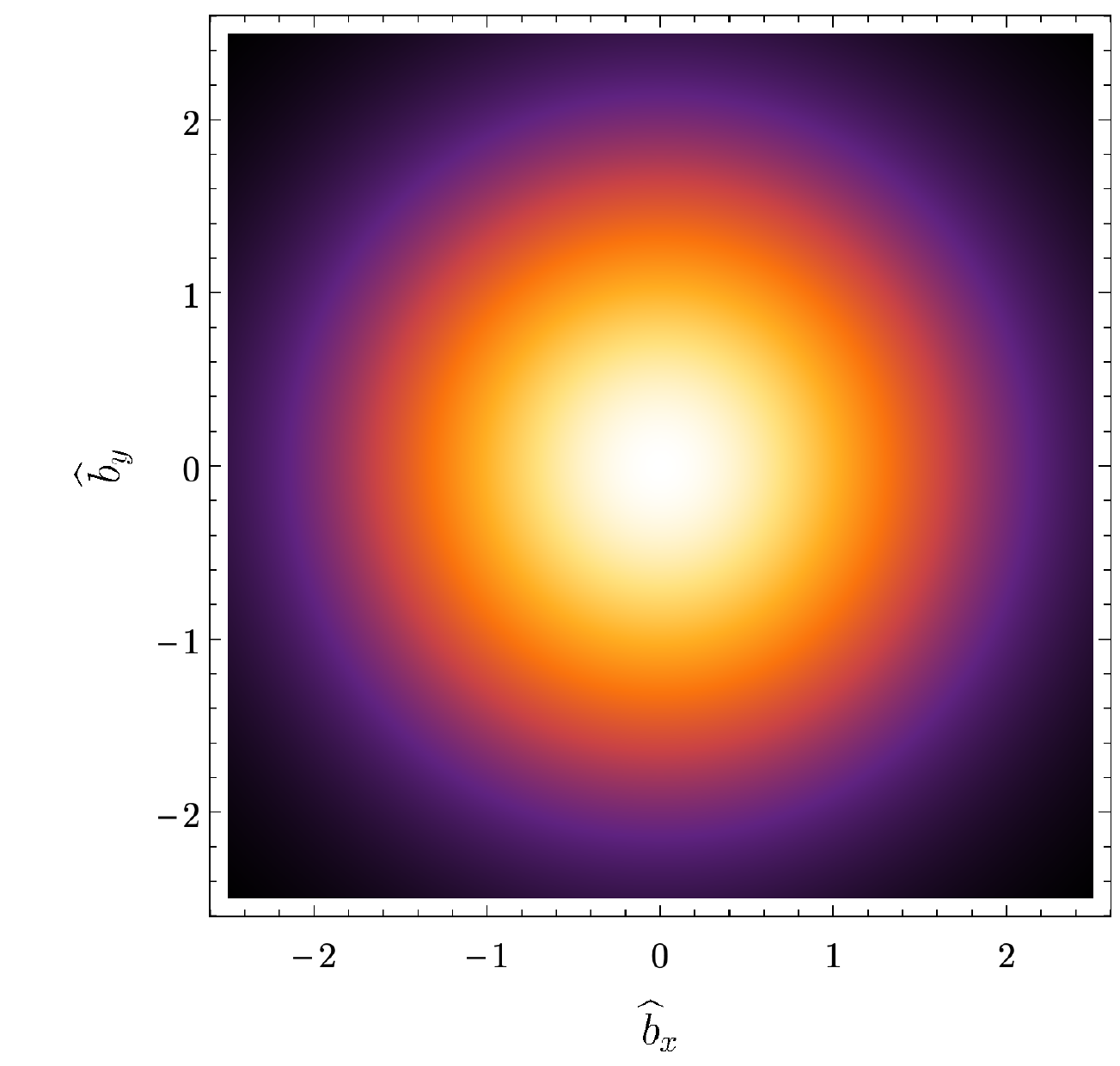}\hspace*{2ex } &
\includegraphics[clip,width=0.425\linewidth]{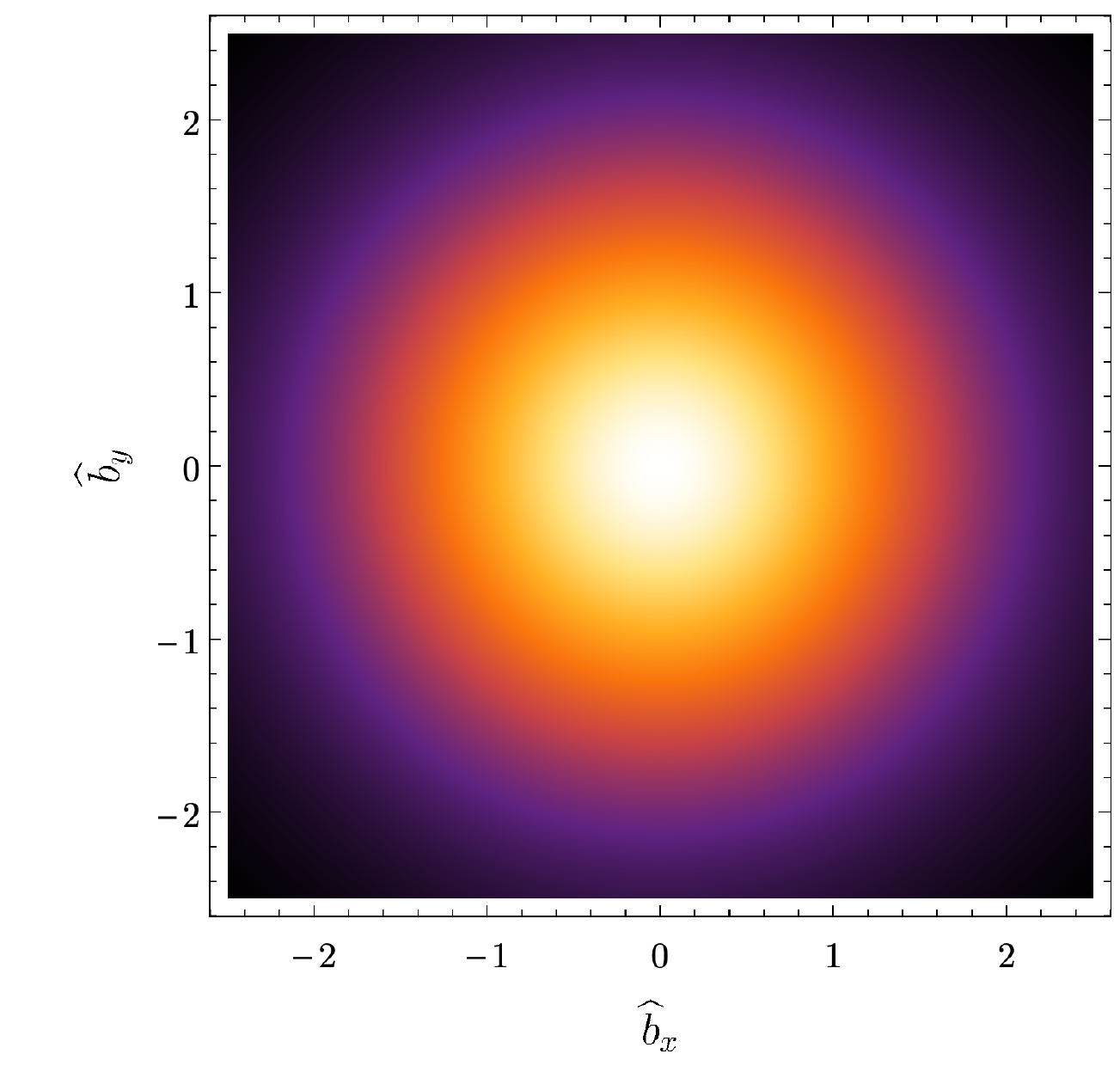}\vspace*{-1ex}
\end{tabular}
\end{center}
%
\caption{\label{FigfdensityF2}
Light-front transverse valence $u$- and $d$-quark anomalous magnetisation densities.
\underline{Upper panels} -- one-dimensional profiles defined in Eq.\,\eqref{fdensityF2}: left,  $\hat\rho_2^f(|\hat b|)$; and right, $|\hat b|\hat\rho_2^f(|\hat b|)$.  The shaded bands centred on each curve mark the $1\sigma$ uncertainty for the SPM approximants; and the $d$-quark density is multiplied by ``$-1$'' to simplify visual comparisons.
\underline{Lower panels} -- two dimensional images derived from the curves in the upper-left panel: \emph{left}, $\hat\rho_2^u(|\hat b|)$; and \emph{right}, $-\hat\rho_2^d(|\hat b|)$.
}
\end{figure*}

\subsubsection{Anomalous magnetisation}
\label{AnomMag}
A valence-quark anomalous magnetisation density can similarly be defined:
\begin{equation}
\label{fdensityF2}
\hat \rho_2^f(|\hat b|) = \int\frac{d^2 \vec{q}_\perp}{(2\pi)^2}\,{\rm e}^{i \vec{q}_\perp \cdot \hat b}
F_2^{f}(Q^2)\,.
\end{equation}
Naturally,
\begin{equation}
\kappa_f = \int d^2\hat b \, \hat \rho_2^f(|\hat b|) \,.
\end{equation}
Our computed values are $\kappa_d=-1.68$, $\kappa_u=1.41$.

We depict the valence $u$- and $d$-quark light-front-transverse anomalous magnetisation densities in Fig.\,\ref{FigfdensityF2}.  The images yield a fair amount of information.

(\emph{i}) The array of panels in Fig.\,\ref{FigfdensityF2} reveal that the valence $d$-quark is magnetically very active within the proton: even though there are two $u$-quarks and only one $d$-quark, the $d$-quark profiles are approximately equal in magnitude (opposite in sign) to the $u$-quark profiles, which are not here divided by two.  This was remarked upon in connection with Fig.\,\ref{FlavourSeparated}; and the explanation can be found by studying Ref.\,\cite[Fig.\,5]{Cloet:2008re}.
The contribution from the two $u$-quarks is mainly generated by the scalar-diquark piece of Diagram~1 in Fig.\,\ref{vertexB}.  All other terms interfere destructively.  Since scalar diquarks are magnetically inert, this term is the weakest source of magnetisation current.
On the other hand, the $d$-quark magnetisation contribution arises primarily from those diagrams with a pseudovector diquark in both the initial- and final-state proton, all of which involve maximal spin and/or angular momentum.  Moreover, the photon-quark and photon-diquark scattering terms interfere constructively.
These remarks highlight that if the proton did not contain a pseudovector diquark component, then the picture would change dramatically, \emph{viz}.\ the $u$-quarks' contribution to the magnetisation would be far greater than that from the $d$-quark.

(\emph{ii}) These conclusions are supported by a comparison between the $u$- and $d$-quark anomalous magnetisation radii, defined in analogy with Eq.\,\eqref{TransverseRadii}:
\begin{equation}
\hat \lambda_2^u \approx 2.9 \,,\; \hat \lambda_2^d =3.7\,.
\end{equation}
Using the parametrisations of data in Ref.\,\cite{Kelly:2004hm}, one obtains the following values, respectively: $3.0$, $3.7$.  Evidently, the anomalous magnetisation density connected with the $d$-quark spreads further from the proton's CoTM than that associated with any given valence $u$-quark.  This conclusion is visually supported by all panels of Fig.\,\ref{FigfdensityF2}.

\subsubsection{Observations}
In closing this section, we reiterate that the Faddeev equation in Fig.\,\ref{figFaddeev} describes the nucleon's dressed-quark core.  Hence, all statements made herein describe that part of a nucleon.  Meson cloud effects may be expected to contribute on some bounded domain beyond $|\hat b| \gtrsim 4$ ($\approx 0.8\,$fm), adding qualitatively to features of these distributions at such larger separations from the nucleon's CoTM.

It is also worth highlighting that the nucleon ground state is just one isolated system in a rich spectrum of baryons; consequently, even a complete explanation of nucleon properties reveals only a small part of a larger picture.  A greater depth of field is achieved by also studying nucleon-to-resonance transitions \cite{Mokeev:2015moa, Burkert:2019opk}.  Here, too, transverse densities can be valuable in developing insights into strong QCD \cite{Carlson:2007xd, Tiator:2008kd, Roberts:2018hpf}.

Finally, note that since the elastic and transition form factors are Poincar\'e-invariant and empirically observable, then all associated remarks and conclusions are independent of renormalisation scale and an observer's frame of reference.  Concerning the light-front-transverse densities, one is treating a particular projection of Poincar\'e-invariant functions; hence, they are frame specific, not frame dependent.

\section{Summary and Outlook}
\label{Epilogue}
Using a Poincar\'e-covariant quark+diquark Faddeev equation for the nucleon, we delivered predictions for all nucleon elastic form factors on $0\leq Q^2\leq 18 m_N^2$, a domain that is expected to be covered by forthcoming experiments at JLab.  The Faddeev equation is defined by a kernel whose structure is informed by nonperturbative results for QCD's Schwinger functions and therefore incorporates important dynamical features, such as dressed-quark mass-functions and momentum-dependent amplitudes for the diquark correlations, whose origin is tied to the emergence of hadronic mass (EHM) within the Standard Model \cite{Roberts:2019ngp}.

The calculated form factors match well with available data, which in some instances extend to $Q^2 \approx 10\,m_N^2$.  In addition to assisting with the effective conduct of new generation experiments, the predicted behaviour of the form factors on $Q^2\gtrsim 10\,m_N^2$ exposes many empirically verifiable features.  Notable amongst them are: a zero in $G_E^p/G_M^p$ and a maximum in $G_E^n/G_M^n$ (Fig.\,\ref{GEonGM}); and a zero in the proton's $d$-quark Dirac form factor, $F_1^d$ (Fig.\,\ref{FlavourSeparated}).  Additional novel predictions (Secs.\,\ref{SecElastic}, \ref{SecFlavour}) lie beyond the range accessible to the 12\,GeV JLab, but could be tested with a high-luminosity, higher-energy accelerator.

Interesting, too, are our predictions for the flavour-separated light-front-transverse number and anomalous magnetisation densities (Sec.\,\ref{SubSecLFTDs}), which deliver some valuable insights.  For example, identifying $|\hat b|$ as the transverse distance from the proton's centre of transverse momentum: there is an excess of valence $u$-quarks on $|\hat b|\simeq 0$ and the valence $d$-quark is significantly more active magnetically than either of the valence $u$-quarks.  These features are manifestations of the presence and relative importance of scalar and pseudovector diquark correlations within the proton, which themselves are dynamical consequences of EHM.

Herein, so as to reach large $Q^2$, we employed a new method for interpolating and extrapolating smooth functions (Sec.\,\ref{CalcFFs}): no functional form is assumed; rather, one develops a set of continued fraction interpolations whose extrapolation comes with a robust uncertainty estimate.  The method is applicable to a wide range of problems.  Hence, a natural next step is to use the more fundamental three-quark Faddeev equation of Refs.\,\cite{Eichmann:2011vu, Wang:2018kto, Qin:2019hgk} as the basis for predicting the large $Q^2$ behaviour of nucleon elastic and transition form factors.  Such analysis would both complement and test the quark+diquark picture we have described; and, furthermore, provide a tighter connection between measurable nucleon properties and EHM.

\begin{acknowledgments}
%
%
We are grateful for constructive comments from Y.~Lu, B.~Wojtsekhowski, Y.-Z.~Xu, Z.-N.~Xu and Z.-Q.~Yao;
for the hospitality and support of RWTH Aachen University, III.\,Physikalisches Institut B, Aachen - Germany;
and for access to the C3UPO computer facilities at the Universidad Pablo de Olavide.
Work supported by:
National Natural Science Foundation of China, under grant no.\ 11805097;
Jiangsu Province Natural Science Foundation, under grant no.\ BK20180323;
Jiangsu Province \emph{Hundred Talents Plan for Professionals};
Helmholtz International Center for FAIR, within the LOEWE program of the State of Hesse;
Deutsche Forschungsgemeinschaft, under grant no.\ FI 970/11-1;
and Spanish Ministerio de Econom{\'{\i}}a, Industria y Competitividad (MINECO) under grant no. FPA2017-86380-P.
\end{acknowledgments}



\end{document}